\PassOptionsToPackage{dvipsnames}{xcolor}
\documentclass[twocolumn,twcolappendix]{aastex7}

\usepackage{soul}
\usepackage{hyperref}
\usepackage{mathrsfs}
\usepackage{bm}
\usepackage{times}

\newcommand{\ky}[1]{{#1}}

\shorttitle{SPIn4D II}
\shortauthors{Yang et al.}

\usepackage{amsmath}
\usepackage{graphicx}
\usepackage{CJK}

\makeatletter
\newcommand*{\rom}[1]{\expandafter\@slowromancap\romannumeral #1@}
\makeatother

\graphicspath{{./}{/}}

\begin{document}
\begin{CJK*}{UTF8}{gbsn}

\title{Spectropolarimetric Inversion in Four Dimensions with Deep Learning (SPIn4D): \\
	\rom{2}. A Physics-Informed Machine Learning Method for 3D Solar \\ Photosphere Reconstruction}

\author[0000-0002-7663-7652]{Kai E. Yang (杨凯)}
\affiliation{Institute for Astronomy, University of Hawai`i at M\=anoa, Pukalani, HI 96768, USA}
\email[show]{yangkai@hawaii.edu}

\author[0000-0003-4043-616X]{Xudong Sun (孙旭东)}
\affiliation{Institute for Astronomy, University of Hawai`i at M\=anoa, Pukalani, HI 96768, USA}
\email{xudongs@hawaii.edu}

\author[0000-0002-8259-8303]{Lucas A. Tarr}
\affiliation{National Solar Observatory, Pukalani, HI 96768, USA}
\email{ltarr@nso.edu}

\author[0000-0002-7290-0863]{Jiayi Liu (刘嘉奕)}
\affiliation{Institute for Astronomy, University of Hawai`i at M\=anoa, Honolulu, HI 96822, USA}
\email{jiayiliu@hawaii.edu}

\author[0000-0002-7354-5461]{Peter Sadowski}
\affiliation{Department of Information and Computer Sciences, University of Hawai`i at M\=anoa, Honolulu, HI 96822, USA}
\email{peter.sadowski@hawaii.edu}

\author[0000-0001-6311-146X]{S. Curt Dodds}
\affiliation{Institute for Astronomy, University of Hawai`i at M\=anoa, Honolulu, HI 96822, USA}
\email{dodds@hawaii.edu}

\author[0000-0001-5850-3119]{Matthias Rempel}
\affiliation{High Altitude Observatory, NSF National Center for Atmospheric Research, Boulder, CO 80307, USA}
\email{rempel@ucar.edu}

\author[0000-0001-5459-2628]{Sarah A. Jaeggli}
\affiliation{National Solar Observatory, Pukalani, HI 96768, USA}
\email{sjaeggli@nso.edu}

\author[0000-0002-7451-9804]{Thomas A. Schad}
\affiliation{National Solar Observatory, Pukalani, HI 96768, USA}
\email{tschad@nso.edu}

\author[0000-0001-5171-9144]{Ian Cunnyngham}
\affiliation{Institute for Astronomy, University of Hawai`i at M\=anoa, Honolulu, HI 96822, USA}
\email{ian@cunnyngham.net}

\author[0000-0001-7217-9749]{Yannik Glaser}
\affiliation{Department of Information and Computer Sciences, University of Hawai`i at M\=anoa, Honolulu, HI 96822, USA}
\email{linneamw@hawaii.edu}

\author[0000-0002-2087-1634]{Linnea Wolniewicz}
\affiliation{Department of Information and Computer Sciences, University of Hawai`i at M\=anoa, Honolulu, HI 96822, USA}
\email{yglaser@hawaii.edu}


\begin{abstract}
Inferring the three-dimensional (3D) solar atmospheric structures from observations is a critical task for advancing our understanding of the magnetic fields and electric currents that drive solar activity. In this work, we introduce a novel, Physics-Informed Machine Learning method to reconstruct the 3D structure of the lower solar atmosphere based on the output of optical depth sampled spectropolarimetric inversions, wherein both the fully disambiguated vector magnetic fields and the geometric height associated with each optical depth are returned simultaneously. Traditional techniques typically resolve the 180-degree azimuthal ambiguity assuming a single layer, either ignoring the intrinsic non-planar physical geometry of constant optical-depth surfaces (e.g., the Wilson depression in sunspots), or correcting the effect as a post-processing step. In contrast, our approach simultaneously maps the optical depths to physical heights, and enforces the divergence-free condition for magnetic fields fully in 3D. Tests on magnetohydrodynamic simulations of quiet Sun, plage, and a sunspot demonstrate that our method reliably recovers the horizontal magnetic field orientation in locations with appreciable magnetic field strength. By coupling the resolutions of the azimuthal ambiguity and the geometric heights problems, we achieve a self-consistent reconstruction of the 3D vector magnetic fields and, by extension, the electric current density and Lorentz force. This physics-constrained, label-free training paradigm is a generalizable, physics-anchored framework that extends across solar magnetic environments while improving the understanding of various solar puzzles.
\end{abstract}


\section{Introduction} \label{sec:intro}

Understanding the three-dimensional (3D) structure of the lower solar atmosphere is crucial for unraveling the mechanisms that drive solar activity. The intricate interplay between magnetic fields, plasma flows, and radiative processes creates a highly dynamic environment that underpins phenomena such as solar flares, coronal mass ejections, and coronal heating. An accurate depiction of the 3D structures will not only advance our theoretical comprehension but also enhance our ability to forecast space weather. However, reconstructing this complex structure from remote-sensing observations is inherently challenging. This is partly owing to the complex physical properties of the atmospheric layers themselves, and partly to the limitations of current observational techniques: e.g., degeneracies between the plasma's physical parameters and the resulting radiation, the need to include scattering and non-equilibrium effects for some spectral lines, or the nonlinear mapping between optical depths and physical heights.

Solar magnetic field inferences are one of the most challenging tasks. The observations are usually made for the lower solar atmosphere, e.g., the photosphere and, more often recently, the chromosphere. The inference technique is usually based on the ``inversion'' of the spectropolarimetric data \citep{Auer1977SoPh...55...47A}. For polarization due to the Zeeman effect, the results will inherit the intrinsic 180-degree ambiguity of the magnetic azimuth angle \citep{Harvey1969PhDT.........3H}, i.e., two possible choices of the transverse field vectors that differ by 180 degrees produce the same spectropolarimetric signal. This introduces uncertainty into downstream analyses, such as the extrapolation of the field into the corona, flare forecasting, data-driving simulations, the determination of electric current, flow computations, energy, and helicity flux \citep{Wiegelmann2021LRSP...18....1W,Bobra2015ApJ...798..135B,Kusano2020Sci...369..587K,Jiang2016NatCo...711522J,Kazachenko2014ApJ...795...17K,Liu2012ApJ...761..105L}. 

Various methods have been proposed for the ``disambiguation task'' (resolving the azimuthal ambiguity) in the past decades. Most operate on a single map assumed to lie on a geometrically flat plane.
For example, the Acute Angle Method and its successors compare the observed magnetic field to a reference field, such as the potential field or a linear force-free field, to determine the orientation of the transverse field \citep{Wang1997SoPh..174..265W,Wang2001SoPh..201..323W,Moon2003SoPh..217...79M}. 
The interactive AZAM algorithm for the Advanced Stokes Polarimeter \citep{Elmore1992SPIE.1746...22E} minimizes the angle between neighbor pixels. 
The first iterative approach is the UH Iterative method \citep{Canfield1993ApJ...411..362C}, which applies the acute-angle method using a potential field as the reference field at first, refines it with a linear force-free model, smooths the azimuthal angle locally, and finally minimizes the current and divergence subject to a transverse-field threshold. Since its initial application, various versions of this method have been developed, and it continues to perform well.
The Non-potential Magnetic Field Calculation \citep[NPFC;][]{Georgoulis2005ApJ...629L..69G} method requires the computation of a proxy electric current density. The first version provides a solution similar to the potential field acute angle method when applied to disk center observations, and the refined version (NPFC2) improves the performance by implementing a parity-free absolute vertical current density \citep{Semel1998A&A...331..383S}. The Minimum Energy method \citep[ME0;][]{Metcalf1994SoPh..155..235M} tries to minimize a function that contains the residual divergence of the magnetic field and the electric current via a simulated annealing process. ME0 is widely used for data pipelines, e.g., for the Helioseismic and Magnetic Imager \citep[HMI;][]{Scherrer2012SoPh..275..207S,Hoeksema2014SoPh..289.3483H,Liu2017SoPh..292...29L} and Hinode Solar Optical Telescope/Spectro-Polarimeter \citep[SOT/SP;][]{Tsuneta2008SoPh..249..167T}. New versions of ME0 improve on the temporal consistency of the solution by considering a time sequence of observations simultaneously \citep{Barnes2018csc..confE.116B}, or consider multiple layers at once \cite[ME0-Z;][]{Barnes2012AAS...22020609B}.
More disambiguation techniques and their performance evaluation can be found in the reviews \citep[][and reference therein]{Metcalf2006SoPh..237..267M,Leka2009SoPh..260...83L,Leka2012SoPh..276..441L}.

Other non-optimization based methods have also been proposed. For example, the ambiguity of the transverse field can be resolved by using multi-viewpoint measurements with at least $0.1$ radians separation \citep{Judge2019MNRAS.482.5542J}. This has been realized \citep{Valori2022SoPh..297...12V,Valori2023A&A...677A..25V} by coordinated observations from \textit{Solar Orbiter} \citep[\textit{SolO};][]{Muller2020A&A...642A...1M} and \textit{Solar Dynamic Observatory} \citep[\textit{SDO};][]{Pesnell2012SoPh..275....3P}. Nevertheless, the number of such joint observations is extremely limited, and their spatial resolution is not sufficient to study small-scale (sub-arcsec) phenomena, such as fine features in quiet-Sun granules or photospheric bright points. 
A second example is a spectroscopic method using the properties of resonance lines under anisotropic radiation fields \citep{Landi1993ApJ...411L..49L}, e.g., the \ion{Sr}{1} 4607 \AA{} line. However, as pointed out by \cite{Judge2019MNRAS.482.5542J}, the conditions required for this are rarely satisfied in the photosphere where observations are typically made, and observing the polarization signal from this line is difficult to begin with \citep{Malherbe2007A&A...462..753M,Bianda2018A&A...614A..89B,Dhara2019A&A...630A..67D,Zeuner2018A&A...619A.179Z,Zeuner2020ApJ...893L..44Z,Zeuner2022A&A...662A..46Z,Zeuner2024ApJ...964...10Z}.

The National Science Foundation's \textit{Daniel K. Inouye Solar Telescope} \citep[DKIST,][]{Rimmele2020SoPh..295..172R} massively expanded the capabilities of ground-based solar observations in multiple ways. Its $4$-m aperture and the excellent seeing conditions at Haleakal\=a allow for high-resolution imaging (down to $0.03\arcsec$), while the off-axis telescope design and optimized development of every optical surface enable precision polarization measurements that far exceed previous capabilities \citep[$\approx 5\times 10^{-4}$;][]{Rimmele2020SoPh..295..172R,deWijn2022SoPh..297...22D,Jaeggli2022SoPh..297..137J,Harrington2023SoPh..298...10H}.


The construction of DKIST coincided with the maturation of sophisticated inversion techniques, whose development over the past decades was driven and continuously supported by Hinode SP observations. With DKIST current operation, these techniques will continue to be advanced and fully exploited to realize its full potential.
Examples include Stokes Inversion based on Response functions \citep[SIR,][]{RuizCobo1992}, Departure coefficient aided SIR \citep[DeSIRe;][]{Cobo2022A&A...660A..37R}, Stokes-Profiles-INversion-O-Routines \citep[SPINOR;][]{Frutiger2000A&A...358.1109F,vanNoort2012A&A...548A...5V}, HAnle and ZEeman Light \citep[HAZEL;][]{AsensioRamos2008ApJ...683..542A}, Non-LTE Inversion COde using the Lorien Engine \citep[NICOLE;][]{Socas-Navarro2015A&A...577A...7S}, SNAPI \citep{Milic2018A&A...617A..24M}, and STockholm Inversion Code \citep[STiC;][]{delaCruzRodriguez2019A&A...623A..74D}. Taking advantage of the high-quality observations, these algorithms can extract information for a range of optical depths where the response function of the spectral line is strong. 

The magnetic fields inferred with the above-mentioned techniques typically lay on surfaces of constant optical depth rather than constant geometric heights. 
This is because they are derived from radiative transfer equations, and the mapping between optical depth and geometric height is not straightforward.
The discrepancy between the two is most apparent in sunspots: the geometric height of a given optical depth is significantly lower to that in the quiet Sun, a phenomenon known as the Wilson depression \citep{Wilson1774RSPT...64....1W}. The strong magnetic field in a sunspot reduces the gas pressure and suppresses convection, leading to a relatively low temperature and low opacity. The same optical height thus corresponds to a physically deeper location in the atmosphere relative to the surrounding quiet sun. This same depression can also be detected for strong magnetic field concentrations in high-resolution, quiet-Sun observations and in plage regions \citep{Kuridze2025ApJ...985L..23K}. All of these phenomena are now readily reproducible in state-of-the-art 3D radiative (magneto-)hydrodynamic solar simulations \citep[see, e.g., Figures 4 and 12 in][]{Yang2024spin4d}. 

Several methods to quantify the Wilson depression have been proposed based on the force balance between pressure gradient, gravity, and Lorentz forces \citep{MartinezPillet1993A&A...270..494M,Mathew2004A&A...422..693M}. They were subsequently extended to include the divergence-free condition of the magnetic field \citep{Puschmann2010ApJ...720.1417P}, which was shown to be effective for addressing the problem \citep{Loptien2018A&A...619A..42L, Loptien2020A&A...635A.202L}.
An alternative approach is to couple the inversion step with the force-balance criteria under the magnetohydrostatic (MHS) approximation, which has been successfully demonstrated using the FIRTEZ code applied to Hinode observations \citep{Yabar2019A&A...629A..24P,Borrero2019A&A...632A.111B,Borrero2021A&A...647A.190B,Borrero2023A&A...669A.122B,Borrero2024A&A...687A.155B}.
A third approach couples the Stokes inversion step directly to a magnetohydro\emph{dynamic} (MHD) simulation, and has been applied to Sunrise/IMaX observation \citep{Riethmuller2017ApJS..229...16R} to provide a full set of atmospheric parameters at each location.
Recently, the Polarimetric and Helioseismic Imager \citep[PHI;][]{Solanki2020A&A...642A..11S} onboard \textit{SolO} provided a direct measurement of the large-scale geometric height using a stereoscopic technique \citep{RomeroAvila2024SoPh..299...41R}.
Most of these methods require a disambiguated vector magnetic field even before assessing the geometric height.

Machine learning (ML) techniques have recently demonstrated their utility in solar physics \citep{AsensioRamos2023LRSP...20....4A}. A notable example is the Stokes Inversion based on Convolutional Neural Networks \citep[SICON;][]{AsensioRamos2019A&A...626A.102A,EstebanPozuelo2024A&A...689A.255E}, which employs convolutional neural networks (CNNs) trained on Max-Planck University-of-Chicago Radiative MHD \citep[MURaM;][]{Vogler2005A&A...429..335V,Rempel2009ApJ...691..640R} sunspot simulations \citep{Rempel2012ApJ...750...62R} and synthesized Stokes profiles. SINCON can directly infer physical and geometric parameters, though it does not address the 180-degree ambiguity. 
Alternative methods have also been proposed that leverage the extensive results from conventional inversion techniques \citep{Higgins2021ApJ...911..130H,Higgins2022ApJS..259...24H}. 
More recently, the Spectropolarimetric Inversion in Four Dimensions with Deep Learning (SPIn4D) project\footnote{\url{https://ifauh.github.io/SPIN4D/}} produced an extensive database of solar atmospheric data of quiet Sun and plage regions from MURaM simulations, encompassing a total of 110 TB of data, and is developing an advanced ML model for analyzing DKIST observed Stokes profiles. An overview of SPIn4D can be found in \cite{Yang2024spin4d}. 

Supervised machine ML models benefit from large volumes of training data generated either by advanced radiative MHD simulations or archival observational results. However, this strength also harbors their limitation: they perform well at interpreting data that are similar to their training data (``interpolation''), but may struggle otherwise (``extrapolation''). If the true solution for a given observation is not well approximated by anything in the database, then the model is unlikely to be able to find it, a problem known as the out-of-distribution issue \citep{hendrycks17baseline}. 
To paraphrase comments at a SHINE workshop in 2022
\begin{quote}
\emph{Models trained by MURaM simulations will only predict MURaM-like solutions.}
\end{quote}
Moreover, neural networks often face interpretability issues \citep{Zeiler10.1007/978-3-319-10590-1_53}, which makes it difficult to predict when a method will fail to generalize.

A promising way to mitigate these issues is to incorporate known physical laws directly into the ML techniques. There are now many such examples
\citep{Weinan2018,SirignanoSIRIGNANO20181339,Lulu2021_19M1274067,Hennigh2021SimNet}, collectively known as physics-informed machine learning \citep[PIML;][and references therein]{Karniadakis2021nature,Kashinath2021RSPTA.37900093K,Hao2022PhysicsInformedML,WU2024124678}. PIML can work with relatively small datasets while ensuring that its predictions obey the prescribed physical laws, typically in the form of partial differential equations (PDEs).
One particular method is the physics-informed neural network \citep[PINN;][]{MaziarRaissiJMLR:v19:18-046,Raissi2019JCoPh.378..686R}.
For specific fluid dynamic problems, this approach proves to achieve an accuracy comparable to traditional computational methods \citep{MAO2020CMAME.360k2789M}. 
PINNs have also been introduced into solar physics problems \ky{\citep{Jarolim2023NatAs...7.1171J,Jarolim2024ApJ...963L..21J,Jarolim2024ApJ...976L..12J,DiazBaso2025A&A...693A.170D}}.
It is worth noting that the PINN architectures are usually used with fixed spatial and temporal coordinates; they do not excel at learning the observational data patterns.


\begin{figure*}
\centering
\fig{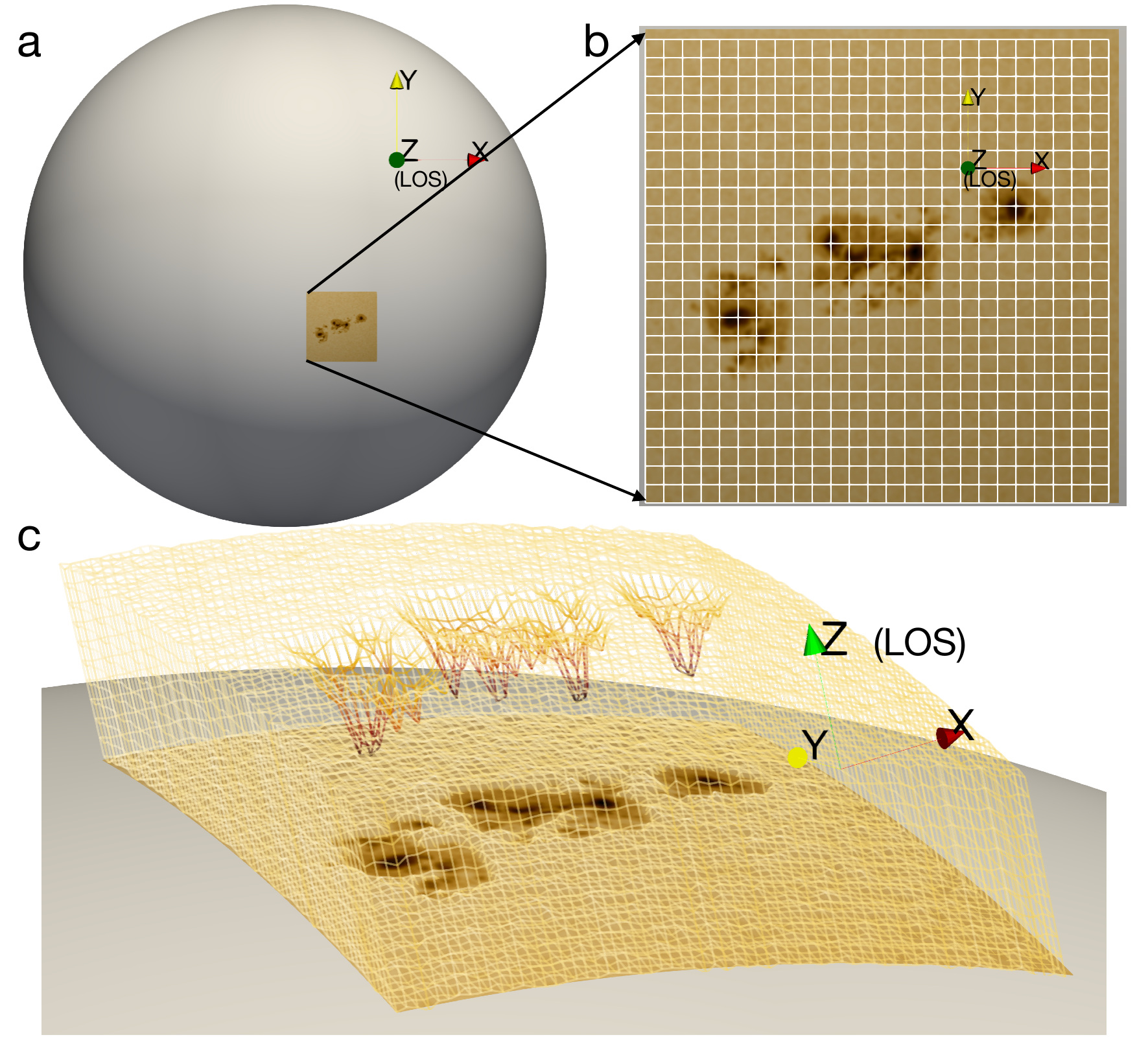}{0.6\textwidth}{}
\caption{An illustration of our task and the mesh grid in our computational domain. Here $x$ and $y$ are image plane axes, and $z$-axis is along the line of sight. The curved mesh presents the depression along the LOS for the optical-depth layers. The example given is NOAA AR 11158 as observed on 2011-02-15 in the continuum by \textit{SDO}/HMI.} \label{fig:geometry}
\end{figure*}


In this paper, as the second paper of the SPIn4D project, we propose a novel method, \texttt{Haleakala Disambiguation Decoder} (HDD), based on the PIML technique, that can address the 180-degree ambiguity of the transverse field and the associated geometric height from multilayer inversion at the same time.
Here, we choose to use the UNet model \citep{Unetpaper}, which is widely used for image segmentation and translation, and is integral to generative artificial intelligence models \citep{Isola8100115,ho2020denoising}. Though simple, it proves to be well-suited for our goals.

The inputs to the method are the magnetic vector components derived from conventional Stokes profiles spanning 3D---two spatial dimensions for the field-of-view (FOV), and one for the optical depth. We thus utilize the 3D UNet model, UNet3D \citep{10.1007/978-3-319-46723-8_49,Wolny10.7554/eLife.57613}, combined with physics-informed constraints, in particular the divergence-free condition. The PIML approach does not rely on any specific numerical simulation model, and can be applied to any inversion results. In the current paper we provide a demonstration of the technique, and apply it only to synthetic observations derived from MURaM simulation data. This allows a precise assessment of the capability of HDD method. The code is openly accessible on GitHub\footnote{\url{https://github.com/Kai-E-Yang/HDD}} \ky{and is also available on Zenodo\footnote{\dataset[doi:10.5281/zenodo.17180997]{https://doi.org/10.5281/zenodo.17180997}.}.}


The paper is organized as follows. \autoref{sec:method} presents the design of the network and the training process. \autoref{sec:test} presents various tests of this technique and a comparison with a conventional method. \autoref{sec:application} discusses two examples of scientific application of our method. Sections \ref{sec:discussion} and \ref{sec:summary} present the discussion and summary, respectively.


\begin{figure*}
\centering
\fig{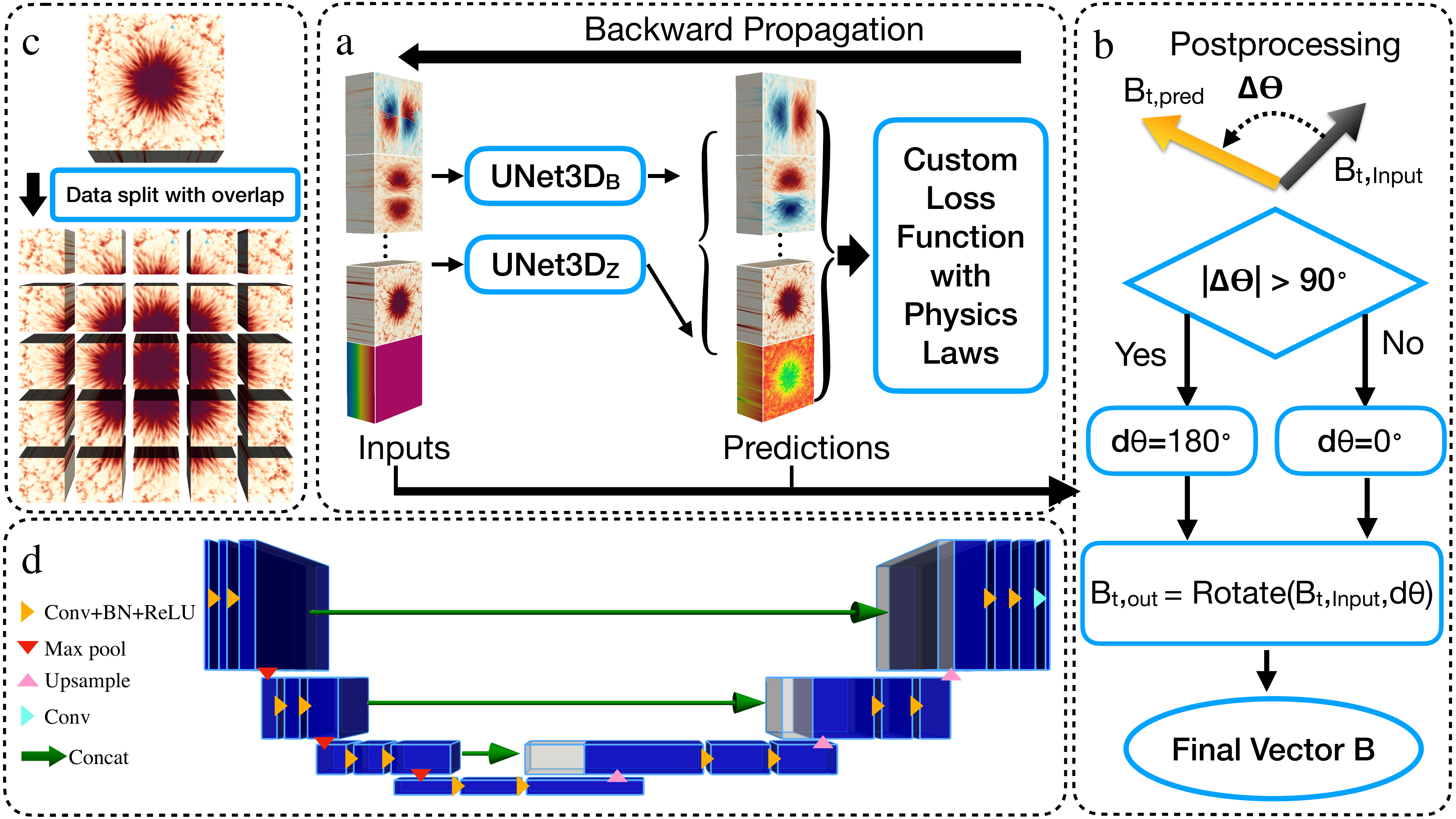}{0.85\textwidth}{}
\caption{Panel (a) shows the flowchart for training the model. The inputs are the 3D magnetic vector fields with azimuthal ambiguity, and an initial guess of the geometric heights associated with each optical-depth layer. The UNet3D$_B$ network predicts 3D vector magnetic field, and the UNet3D$_Z$ network provides the geometric heights. They are combined to compute a custom loss function with physics law encoded (\autoref{eq:loss}). Panel (b) shows the post-processing step that assembles the outputs from the neural networks (prediction) and the original input magnetic fields into the final output. Panel (c) shows an example of the 3D data, split with overlaps between nearby subfields. These individual subfields are used as the input for training. Panel (d) presents the UNet3D structure, reproduced from Figure 2 in \cite{10.1007/978-3-319-46723-8_49}. \label{fig:unet}}
\end{figure*}


\section{Method}\label{sec:method}



An illustration of our tasks is presented in \autoref{fig:geometry}. We target magnetic fields inferred from spectropolarimetric observations of a region on the Sun, for example, an active region (AR; panel b), positioned at some arbitrary location (panel a). Our goal is to reconstruct the magnetic vectors, free from azimuthal ambiguity, on a geometric grid (panel c).

\autoref{fig:unet}(a) provides an overview of the training step. For the input, we start with the magnetic field vectors from inversion (with azimuthal ambiguity) and an initial guess of the geometric height (for each optical depth), both defined on a mesh with a dimension of $(N_x, N_y, N_\tau)$. Here the $x$ and $y$ coordinates span the field of view (FOV), and the $\tau$ coordinate represents the optical depth along the line of sight (LOS). Two individual networks are trained to output the disambiguated magnetic field vectors and the geometric heights (predictions), respectively, on a mesh with the same dimension as the input. \autoref{fig:unet}(b) illustrates a post-processing step that assembles the outputs from the neural networks (prediction) and the input magnetic fields into the final output, i.e., the input magnetic field transverse components from the inversion will be rotated by 180 degrees based on their difference from the network prediction, and this disambiguated magnetic field is combined with the determined geometric height for each grid vertex to comprise the final output. In order to keep the problem computationally tractable, we typically break the input into subdomains with a smaller size $(n_x, n_y, n_\tau)$ as shown in \autoref{fig:unet}(c), which can be recombined at a later stage.

The physical laws we enforce are encoded in the set of custom loss functions used to train each network, including divergence-freeness and monotonicity of the geometric heights. The training of the neural networks is essentially a minimization process of the loss functions that guides the model toward physically meaningful solutions. With two networks trained simultaneously, we effectively address the azimuthal ambiguity and the depression of $\tau$ surfaces (e.g.~sunspot, light bridge, and intergranular lanes) at the same time.

The following subsections describe the coordinate system, network architecture (including the loss functions and the post-processing procedures), and network training strategy, respectively.


\subsection{Coordinate System}\label{sec:coordinates}

We choose a single Cartesian coordinate system for all analysis steps described in this paper. In our setup, we use the right-handed coordinate, the $\hat{\bm{z}}$-axis is aligned with the photon path direction (LOS), and the $\hat{\bm{x}}$ and $\hat{\bm{y}}$ axes correspond to the two transverse observational FOV directions, respectively. In contrast to traditional techniques \citep[e.g.,][]{Metcalf1994SoPh..155..235M} that often perform an explicit transformation into a heliographic frame where the local $\hat{\bm{z}}$ is parallel to gravity \citep{Gary1990SoPh..126...21G}, our method maintains a single coordinate system defined in terms of the LOS direction and the image plane, even for cases that are away from the disc center. This is also the preferred coordinate of most spectropolarimetric inversion algorithms, which provide magnetic-field information at various optical depths along each LOS in the image plane. 
By predicting the geometric height along our $\hat{\bm{z}}$ axis, $Z=Z(\tau)$ for each plane-of-sky location $(X,Y)$, we can conveniently solve the problem for each LOS without the need for additional coordinate transformations. Hereafter, we define the ``vertical'' direction as the $\hat{\bm{z}}$-axis, and the ``transverse" components as those in the $\hat{\bm{x}}$-$\hat{\bm{y}}$ plane, with the transverse magnetic field $B_t=\sqrt{B_x^2+B_y^2}$.

Our coordinate system is equivalent to the ``image coordinates'' $(\xi,\eta,\zeta)$ discussed in \citet{Gary1990SoPh..126...21G}. For a region centered at a heliocentric angle $\theta$ (angle between the LOS and the local normal vector), a plane of constant $Z$ will not be parallel to the solar surface unless $\theta=0$. The $\hat{\bm{z}}$-axis will differ from the local normal, or the direction of the solar gravity $\bm{g}$. As such, a vector transformation will be necessary for analyses conducted in the heliographic frame.


A key feature of our method is the flexible, self-adaptive geometry of the computational domain. While the $\hat{\bm{x}}$ and $\hat{\bm{y}}$ axes remain fixed, the domain is dynamically reshaped along the $\hat{\bm{z}}$ direction to capture the non-planar geometry of the $\tau$ surfaces to accommodate, e.g., the Wilson depression in sunspots. The geometric heights predicted by our neural network will result in a irregular mesh that encodes the true physical scales $Z$ along the LOS (in contrast to $\tau$). Our strategy enables the evaluation of the true spatial derivatives in physical space, which proves to be crucial for ensuring the divergence-free condition $\nabla\cdot \bm{B}=0$ without additional coordinate transformations. A cartoon example of the mesh grid is present in \autoref{fig:geometry}(c). More details on the implementation are presented in Section~\ref{sec:model}.

We note that an observational FOV typically contains several hundred to thousands of pixels along the transverse axes. Spectropolarimetric inversion typically return significantly fewer grid points along the LOS, usually only a few dozen. The number is limited by the physics of formation for each spectral line, the observation's spectral resolution, signal-to-noise level, and the accuracy of the inversion model itself.

Finally, the present work is designed to combine several types of analysis that live naturally on different types of grid with different naming conventions.  
For clarity, we describe them explicitly here.
Observations are usually best described in terms of 2D \emph{pixels} corresponding to, e.g., elements of CCDs, and a $(X,Y)$ coordinate refers to the center of the pixel.
Typical 1D inversions, then reference \emph{nodes} along a line of sight above a 2D pixel, where the vertical spacing of the node is typically defined in terms of optical depth.  Magnetic field values are determined at each node.
Depending on the type of inversion, these may correspond to an average flux through the associated observational pixel (e.g., when a filling factor is used).
Last, we have the coordinates for our neural network, which we describe in detail below, but briefly, it relies on a magnetic field specified at vertices of 3D cells, where the vertices start co-located with the nodes (vertically) and pixel centers (horizontally).


\begin{figure*}
\centering
\fig{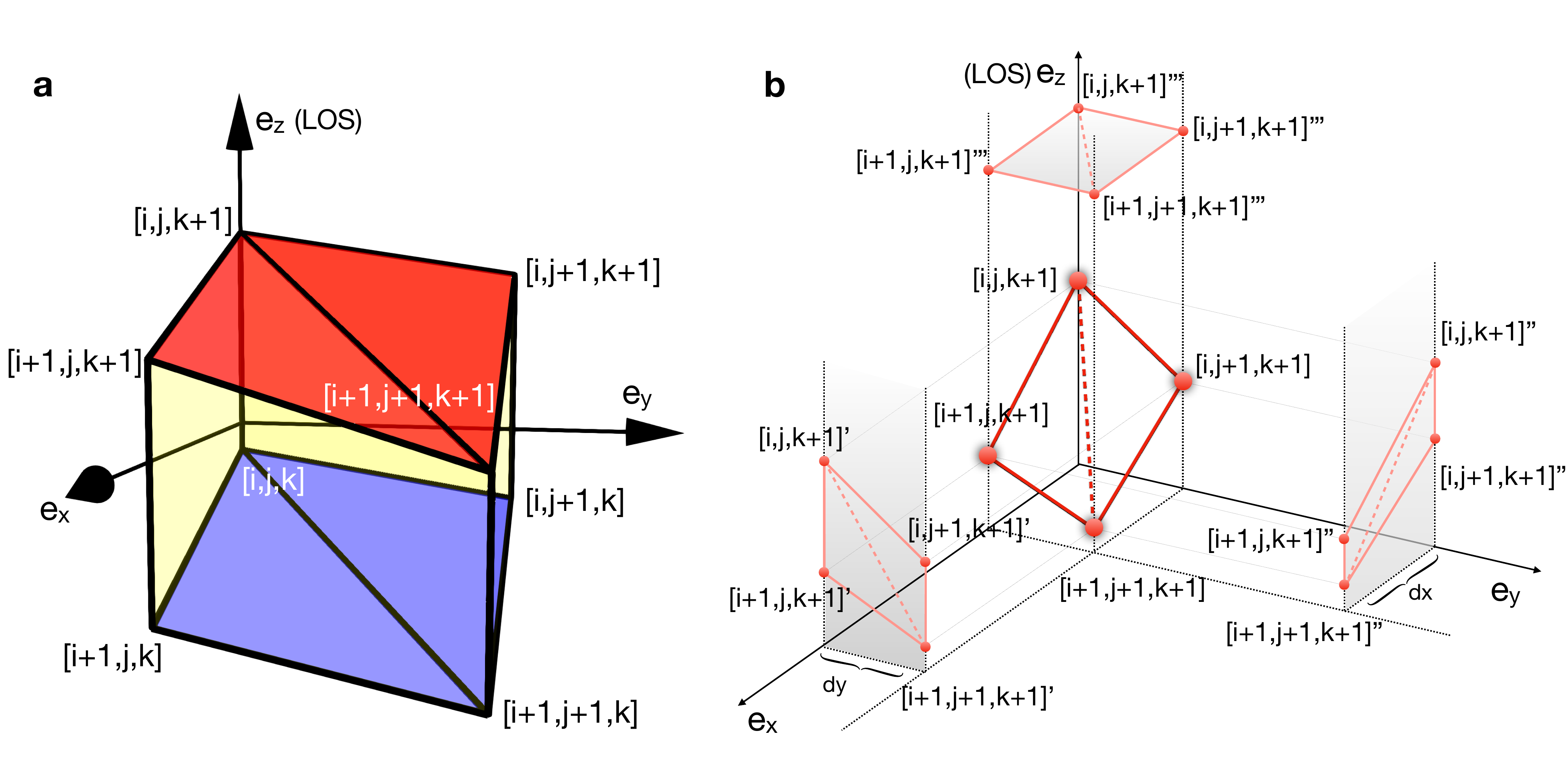}{0.9\textwidth}{}
\caption{Panel (a) illustrates the mesh cell structure and the convention for vertex indexing. Panel (b) shows the top surfaces of a mesh cell at the center, and its projection to three planes perpendicular to $\hat{\bm{x}}$, $\hat{\bm{y}}$, and $\hat{\bm{z}}$, respectively. \label{fig:1}}
\end{figure*}


\subsection{Model and Encoding the Physics}\label{sec:model}


Our network architecture integrates two UNet3D models based on the \texttt{ResidualUNet3D} module \citep{Lee2017SuperhumanAO} from the \texttt{pytorch-3dunet} package \citep{Wolny10.7554/eLife.57613}. An example architecture is shown in \autoref{fig:unet}(d). Each model consists of 4 layers with feature maps of 64, 128, 256, and 512 channels, respectively. Hereafter, ``UNet3D" refers to the \texttt{ResidualUNet3D} module. As illustrated in \autoref{fig:unet}(a), our first model, UNet3D$_{B}$, is tasked to predict a grid of vector magnetic fields with resolved azimuthal angles. The second model, UNet3D$_{Z}$, predicts geometric heights associated with each optical depth.

The input data to these models are a five-dimensional tensor, \ky{i.e., $\mathbf{U}\in\mathbb{R}^{n_b\times n_c\times n_x\times n_y\times n_\tau}$, where $n_b$ and $n_c$ are the batch size and number of channels, the $n_x$, $n_y$, $n_\tau$ indicate the spatial size of the subdomain mentioned above.} 
Specific details regarding the batch size are provided in \autoref{sec:test}.
We define the channel size to be six to accommodate the components of the raw magnetic fields (with azimuthal ambiguity) and an initial estimate of the geometric height. One channel refers to the vertical magnetic field, and four channels refer to the transverse field components. We note that the latter consists of two copies of the same $(B_x, B_y)$ with different realizations of the azimuthal ambiguity: one copy has its azimuth modulated into the range $[ 0,\pi ]$, while the other into the range $[-\pi/2,\pi/2 ]$. Such redundancy appears to improve the model accuracy. All magnetic field components used here are normalized by the maximum magnitude value across the entire dataset. The last channel of the input is the geometric height; the initial estimates are defined as a uniform distribution with a constant thickness for each $\Delta\log_{10}\tau=0.1$, based on the median value from the MURaM simulation. The final three \ky{axes of the input tensor, ($n_x$, $n_y$, $n_\tau$),} represent spatial coordinates $X_{ijk}$, $Y_{ijk}$, and $Z_{ijk}$, spanning the field of view (FOV) and optical depth layers accordingly. Here the $i$, $j$, and $k$ subscripts denote the vertex indexes along $\hat{\bm{x}}$, $\hat{\bm{y}}$, and $\hat{\bm{z}}$ directions in our defined image-plane coordinate system (see Section \ref{sec:coordinates}), defined at cell vertices. \autoref{fig:1} illustrates the cell structure of this grid.

The outputs from UNet3D$_{B}$ and UNet3D$_{Z}$ consist of three channels for the magnetic field and one channel for the geometric height, respectively. Thus, each batch of training data yields predictions for both the magnetic field $\bm{B}_{ijk} = (B_{x,ijk}, B_{y,ijk}, B_{z,ijk})$ and geometric height $Z_{ijk}$, 
Note that the $\hat{\bm{z}}$-direction corresponds to the LOS and is the same as $\tau$ orientation. As a result, the predicted vector magnetic field and geometric height values are defined at each pixel in the FOV across all optical depth layers.

We note that for a region away from the disc center, at a heliocentric angle of $\theta$, the $k$-th layer will initially lie in a plane with constant $Z$, and will form an angle of $\theta$ with the solar surface. As the training progresses, the average direction of this layer should gradually change and becomes largely parallel to the solar surface, as expected for a layer with constant optical depth. For this work, we focus on the case with $\theta=0$.

Our algorithm requires a post-processing step to assemble the network-predicted field $\bm{B}_{\rm pred}$ and the input field $\bm{B}_{\rm in}$ into the final output, as illustrated in \autoref{fig:unet}(b). This step is necessary because the neural networks only provide a $\bm{B}_{\rm pred}$ solution that best satisfies our physics-informed constraint. There is no explicit feedback to forward synthesis of the spectral lines, and the transverse components are weakly constrained to have the same magnitude and (anti-)parallel to the spectrally best-fit $\bm{B}_{\rm in}$. Therefore, to finally resolve the azimuthal ambiguity in $\bm{B}_{\rm in}$, we use the network prediction as a reference. Specifically, when the absolute angle difference between the input field and the neural network prediction is greater than 90 degree, we flip the orientation of the input magnetic field. This step also proves to reduce artifacts (e.g., checkerboard patterns discussed in \autoref{sec:test}) that can arise from imperfect neural network training. The resulting, disambiguated field is then well suited for further scientific analysis, such as estimating the electric currents and the Lorentz force.

Our custom loss functions encode the physical constraints, including the difference between the predicted magnetic field $\bm{B}_{\rm pred}$ and the input magnetic field $\bm{B}_{\rm in}$ components, as well as the divergence of $\bm{B}_{\rm pred}$. Specifically, the loss functions for the magnitude of the LOS and transverse field are
\begin{equation}\label{eq:bz}
    {\rm loss}_{B_z}= \left< \frac{(B_{z,{\rm pred}}-B_{z,{\rm in}})^4}{B_{z,{\rm in}}^2+\epsilon} \right>,
\end{equation}
and
\begin{equation}\label{eq:bt}
    {\rm loss}_{B_t}= \left< \frac{(B_{x,{\rm pred}}^2+B_{y,{\rm pred}}^2 - B_{x,{\rm in}}^2-B_{y,{\rm in}}^2)^2}{B_{x,{\rm in}}^2+B_{y,{\rm in}}^2+\epsilon} \right>,
\end{equation}
respectively, where the angle brackets denote the mean values computed over all vertices. The predicted transverse field should also be close to (anti-)parallel to the input field. This property is enforced by minimizing the square of the cross-product between the two transverse fields, that is,
\begin{equation}\label{eq:parallel}
{\rm loss}_{\rm parallel}= \left< \frac{(B_{x,{\rm pred}}B_{y,{\rm in}} - B_{y,{\rm pred}}B_{x,{\rm in}})^2}{|B_{\rm in}|^2+\epsilon} \right>.
\end{equation}
For all loss functions, a small regularization term $\epsilon=10^{-8}$ is added to the denominator to avoid division by zero.
\ky{We note that in Equation (\ref{eq:bz})--(\ref{eq:parallel}), the numerators and denominators use a fourth and second power, respectively. These exponents are empirically determined through extensive experimentation.}

Our most important constraint, the solenoidal condition for magnetic fields, is implemented using a finite-volume-like approach. That is, we integrate $\nabla\cdot\bm{B}=0$ over each cell and define the relevant loss function in terms of the net magnetic flux residuals across each grid cell in the predicted field. This step requires extra care to account for the self-adapting, deformable grid, which we now describe in detail.

At each vertex address $(i,j,k)$, the networks predict a vector magnetic field $\bm{B}_{ijk}$ and a geometric height $Z_{ijk}$. The predicted dataset has the same dimensions as the input. The transverse coordinates are fixed and given by $X_{ijk} = i\Delta x + X_{\text{min}}$ and $Y_{ijk} = j \Delta y + Y_{\text{min}}$, where $\Delta x$ and $\Delta y$ refer to the spatial step size in $\hat{\bm{x}}$ and $\hat{\bm{y}}$ directions, respectively. Here $X_{\text{min}}$ and $Y_{\text{min}}$ are the minima of the transverse coordinate, which are used as reference. The LOS coordinate $Z_{ijk}$ is allowed to vary during the network training. We can define a mesh cell consisting of eight ``nodes'' (vertices): $\vec{r}_{i,j,k}$, $\vec{r}_{i+1,j,k}$, $\vec{r}_{i,j+1,k}$, $\vec{r}_{i+1,j+1,k}$, $\vec{r}_{i,j,k+1}$, $\vec{r}_{i+1,j,k+1}$, $\vec{r}_{i,j+1,k+1}$, and $\vec{r}_{i+1,j+1,k+1}$, where $\vec{r}_{i,j,k}=(X_{ijk}, Y_{ijk}, Z_{ijk})$ indicates the vertex coordinate vector, as illustrated in \autoref{fig:1}. Such a setup allows us to calculate the net magnetic flux across all surfaces of each cell, and the mean value of these net fluxes over the entire dataset serves as a measure of the divergence-free quality. This mean flux is added to the custom loss function, and will be minimized by the neural networks training along with other terms.

The surfaces of each cell shown in \autoref{fig:1}(a) can be classified into three groups: lateral surfaces (transparent yellow), top surfaces (red), and bottom surfaces (blue). The lateral surfaces are defined as
\begin{equation}
\begin{split}
S_{l,1} &= (\vec{r}_{i,j,k}, \vec{r}_{i+1,j,k}, \vec{r}_{i+1,j,k+1}, \vec{r}_{i,j,k+1}),\\ 
S_{l,2} &= (\vec{r}_{i+1,j,k}, \vec{r}_{i+1,j+1,k}, \vec{r}_{i+1,j+1,k+1}, \vec{r}_{i+1,j,k+1}),\\ 
S_{l,3} &= (\vec{r}_{i+1,j+1,k}, \vec{r}_{i,j+1,k}, \vec{r}_{i,j+1,k+1}, \vec{r}_{i+1,j+1,k+1}),\\ 
S_{l,4} &= (\vec{r}_{i,j+1,k}, \vec{r}_{i,j,k}, \vec{r}_{i,j,k+1}, \vec{r}_{i,j+1,k+1}).
\end{split}
\end{equation}
The top surfaces are defined as
\begin{equation}
\begin{split}
S_{{\rm top},1} &= (\vec{r}_{i,j,k+1},\vec{r}_{i+1,j,k+1},\vec{r}_{i+1,j+1,k+1}),\\
S_{{\rm top},2} &= (\vec{r}_{i,j,k+1},\vec{r}_{i+1,j+1,k+1},\vec{r}_{i,j+1,k+1}).
\end{split}
\end{equation}
The bottom surfaces are defined as
\begin{equation}
\begin{split}
S_{{\rm bottom},1} =& (\vec{r}_{i,j,k},\vec{r}_{i+1,j,k},\vec{r}_{i+1,j+1,k}),\\
S_{{\rm bottom},2} =& (\vec{r}_{i,j,k},\vec{r}_{i+1,j+1,k},\vec{r}_{i,j+1,k}).
\end{split}
\end{equation}
Because the $X$ and $Y$ coordinates remain unchanged during training, the four nodes of each lateral surface are always coplanar and perpendicular to the $\hat{\bm{z}}$ direction. The top (bottom) surfaces are broken into two triangle simplices, as the four nodes with subscript $k+1$ ($k$) are not necessarily coplanar after training.

The area of the four lateral surfaces, $A(\text{surface})$, can be easily computed as 
\begin{equation}
\begin{split}
A(S_{l,1}) = 0.5\Delta x &(Z_{i,j,k+1}-Z_{i,j,k}\\
&+Z_{i+1,j,k+1}-Z_{i+1,j,k}),\\
A(S_{l,2}) = 0.5\Delta y &(Z_{i+1,j,k+1}-Z_{i+1,j,k}\\
&+Z_{i+1,j+1,k+1}-Z_{i+1,j+1,k}),\\
A(S_{l,3}) = 0.5\Delta x &(Z_{i+1,j+1,k+1}-Z_{i+1,j+1,k}\\
&+Z_{i,j+1,k+1}-Z_{i,j+1,k}),\\
A(S_{l,4}) = 0.5\Delta y &(Z_{i,j+1,k+1}-Z_{i,j+1,k}\\
&+Z_{i,j,k+1}-Z_{i,j,k}).
\end{split}
\end{equation}
The magnetic fluxes across these four lateral surfaces are 
\begin{equation}
\begin{split}
{\rm Flux}(S_{l,1})=0.25&A(S_{l,1})\times(B_{y,i,j,k}+B_{y,i+1,j,k}\\
&+B_{y,i+1,j,k+1}+B_{y,i,j,k+1}),\\
{\rm Flux}(S_{l,2})=0.25&A(S_{l,2})\times(B_{x,i+1,j,k}+B_{x,i+1,j+1,k}\\
&+B_{x,i+1,j+1,k+1}+B_{x,i+1,j,k+1}),\\
{\rm Flux}(S_{l,3})=0.25&A(S_{l,3})\times(B_{y,i+1,j+1,k}+B_{y,i,j+1,k}\\
&+B_{y,i,j+1,k+1}+B_{y,i+1,j+1,k+1}),\\
{\rm Flux}(S_{l,4})=0.25&A(S_{l,4})\times(B_{x,i,j+1,k}+B_{x,i,j,k}\\
&+B_{x,i,j,k+1}+B_{x,i,j+1,k+1}).
\end{split}
\end{equation}

Calculating the fluxes across the top and bottom surfaces requires more care because the $Z$ coordinate varies from vertex to vertex. We choose to project the two triangular simplices of the top and bottom surfaces onto planes perpendicular to $\hat{\bm{x}}$, $\hat{\bm{y}}$, and $\hat{\bm{z}}$, rather than projecting the vector magnetic field components. Then the flux through each original surface is computed as the sum of the fluxes from each projected component. An example of the top surfaces and their projection are illustrated in \autoref{fig:1}(b).
The projected areas on the plane with normal vector $\hat{\bm{x}}$ are
\begin{equation}
\begin{split}
A_x(S_{{\rm top},1}) &= 0.5\Delta y(Z'_{i+1,j,k+1}-Z'_{i,j,k+1}),\\
A_x(S_{{\rm top},2}) &= 0.5\Delta y(Z'_{i,j+1,k+1}-Z'_{i+1,j+1,k+1}),
\end{split}
\end{equation}
and those on the plane with normal vector $\hat{\bm{y}}$ are
\begin{equation}
\begin{split}
A_y(S_{{\rm top},1}) &= 0.5\Delta x(Z''_{i+1,j,k+1}-Z''_{i+1,j+1,k+1}),\\
A_y(S_{{\rm top},2}) &= 0.5\Delta x(Z''_{i,j,k+1}-Z''_{i,j+1,k+1}),
\end{split}
\end{equation}
where the superscripts $'$ and $''$ indicate the projected variables. The projected areas on the $\hat{\bm{z}}$-perpendicular plane are
\begin{equation}
\begin{split}
A_z(S_{{\rm top},1}) &= 0.5\Delta x\Delta y,\\
A_z(S_{{\rm top},2}) &= 0.5\Delta x\Delta y.
\end{split}
\end{equation}
Therefore, the fluxes are
\begin{equation}
\begin{split}
{\rm Flux}(S_{{\rm top},1})&=A_x(S_{{\rm top},1})\times (B_{x,i,j,k+1}+\\
&B_{x,i+1,j,k+1}+B_{x,i+1,j+1,k+1})/3\\
&+A_y(S_{{\rm top},1})\times (B_{y,i,j,k+1}+\\
&B_{y,i+1,j,k+1}+B_{y,i+1,j+1,k+1})/3\\
&+A_z(S_{{\rm top},1})\times (B_{z,i,j,k+1}+\\
&B_{z,i+1,j,k+1}+B_{z,i+1,j+1,k+1})/3,\\
{\rm Flux}(S_{{\rm top},2})&=A_x(S_{{\rm top},2})\times (B_{x,i,j,k+1}+\\
&B_{x,i+1,j+1,k+1}+B_{x,i,j+1,k+1})/3\\
&+A_y(S_{{\rm top},2})\times (B_{y,i,j,k+1}+\\
&B_{y,i+1,j+1,k+1}+B_{y,i,j+1,k+1})/3\\
&+A_z(S_{{\rm top},2})\times (B_{z,i,j,k+1}+\\
&B_{z,i+1,j+1,k+1}+B_{z,i,j+1,k+1})/3.
\end{split}
\end{equation}
The fluxes on the bottom surfaces can be computed similarly, but their surface normal orientations are reversed compared to those of the top surfaces.
Finally, we compute the total flux through each cell $n$ as
\begin{equation}\label{eq:flux}
\begin{split}
{\rm Flux}_n=&\bigl[{\rm Flux}(S_{l,1})+{\rm Flux}(S_{l,2})+{\rm Flux}(S_{l,3}) \\
&+{\rm Flux}(S_{l,4}) + {\rm Flux}(S_{{\rm top},1})+ {\rm Flux}(S_{{\rm top},2}) \\
&+{\rm Flux}(S_{{\rm bottom},1})+ {\rm Flux}(S_{{\rm bottom},2})\bigr].
\end{split}
\end{equation}
For a divergence free field, the net flux through each cell is zero, individually.  


\begin{deluxetable}{cc} 
\tablecaption{HDD Weight Values\label{tab:weight}}
\tablewidth{0pt}
\tablehead{
\colhead{Weight} & \colhead{Value}}
\startdata
$w_{B_z}$& $10^8$ \\
$w_{B_t}$& $10^7$ \\
$w_{\rm parallel}$& $10^{10}$ \\
$w_{\rm div}$& $10^8$ \\
$w_{\rm div,0}$\tablenotemark{\textup{*}} & $0.5W\times(\tanh(10\times(\dfrac{{\rm epoch}}{20000}-0.5))+1)+1$ \\
$w_{\rm smooth}$& $10^4$ \\
$w_{\rm mom}$\tablenotemark{\textup{\textdagger}}& $10^{15}$
\enddata
\tablecomments{
\tablenotetext{\text{*}}{We set the constant $W=10^8-1$, such that the weight drops from $10^8$ toward $1$ as training progresses. This term is used as an initial estimate of the residual flux and intended for stabilizing the training.}
\tablenotetext{\text{\textdagger}}{We use a large value to ensure that the geometric height $Z$ increases monotonically.}
}
\vspace{-5mm}
\end{deluxetable}


The final loss function for divergence is
\begin{equation}\label{eq:loss_div}
{\rm loss}_{\rm div} = \left< \frac{{\rm Flux}_n^4}{{\rm avg}(B_{x})^2 + {\rm avg}(B_{y})^2 + {\rm avg}(B_{z})^2 + \epsilon} \right>,
\end{equation}
where the operator ${\rm avg}()$ takes the average of the eight nodes defining each cell, i.e., the vertices in \autoref{fig:1}(a), and the mean is taken over all cells.

We introduce an additional loss function ${\rm loss}_{{\rm div},0}$, which approximates the net residual flux across each cell if the coordinates were fixed, i.e., the case where $Z$ does not change during training. We find such a term to be crucial for training stability during the initial stages. It acts as a first-order estimate of the true divergence loss: it thus provides crucial information to the neural network when it does not know yet how to predict the geometric heights. As the training progresses, we gradually reduce the importance of this term so the true divergence loss term ${\rm loss}_{\rm div}$ takes over, as detailed in Section~\ref{sec:train}. The loss term ${\rm loss}_{{\rm div},0}$ takes a similar form as \autoref{eq:loss_div}, but has the grid size $\Delta Z$ set to a constant value of $14.4$~km, corresponding to the typical thickness of $\Delta\log_{10}\tau=0.1$ in the MURaM simulation. We find that the result is insensitive to the grid size.

Two additional loss terms are required to avoid pathological $Z$ outputs.
First, we add a 3D median filter in the predicted geometric height to keep the predicted 3D geometric smooth (i.e., no large jumps or extreme values), 
\begin{equation}
{\rm loss}_{\rm smooth} = \left< [Z - {\rm median\_filter}(Z)]^2 \right>,
\end{equation}
where we use a 3D kernel size of $5\times 5\times 5$ local neighborhood of vertices.
Second, we require the predicted geometric height to increase monotonically in the $\hat{\bm{z}}$ direction, that is,
\begin{equation}
{\rm loss}_{\rm mon}=\max \{0, -(\delta Z_{i,j,k} - \delta Z_{\rm min})\},
\end{equation}
where $\delta Z_{i,j,k}=Z_{i,j,k+1}-Z_{i,j,k}$ is the thickness between two optical depth layers. The value $\delta Z_{\rm min}=1.1$~km is a predetermined minimum physical distance corresponding to an optical depth difference of $\delta\tau=0.1$ in the MURaM quiet-Sun simulation. Strict monotonicity of the grid can be ensured by setting a large weight for this term in the final loss function.

The final custom loss function is
\begin{equation}\label{eq:loss}
\begin{split}
{\rm loss}&=w_{B_z}{\rm loss}_{B_z} + w_{B_t}{\rm loss}_{B_t}\\
&+w_{\rm parallel}{\rm loss}_{\rm parallel} + w_{\rm div}{\rm loss}_{\rm div}\\
&+w_{\rm div,0}{\rm loss}_{\rm div,0}+w_{\rm smooth}{\rm loss}_{\rm smooth}\\
&+ w_{\rm mon}{\rm loss}_{\rm mon}.
\end{split}
\end{equation}
The weight value for each loss term is summarized in \autoref{tab:weight} and discussed below.



\subsection{Training Strategy}\label{sec:train}


To train the model with the type of magnetic field data typical of results from a solar inversion code, we first partition the input dataset into smaller, overlapping patches. For this work, we use a size of $(n_x,n_y,n_\tau)=(96,96,32)$ pixels. We note that the FOV of a spectropolarimetric dataset typically spans hundreds to several thousand pixels. The number of optical-depth layers provided by inversion are typically less than $32$, but are customarily interpolated to finer $\tau$ grids in the literature for further analysis.

An overlap of 16 pixels is introduced in both $\hat{\bm{x}}$- and $\hat{\bm{y}}$-dimensions between neighboring patches. This strategy insures that important spatial information is preserved across the patch boundaries, so the model can learn patterns with spatial scales larger than the size of a single patch. The scheme is widely used when training UNet models \citep{Unetpaper,10.1007/978-3-319-46723-8_49,nn-Unet}.
The overlapping patches are then aggregated into batches for network training, which enhances the model's ability to capture spatial variations in the magnetic field and mitigates the edge artifacts from data padding in Unet3D. This process is illustrated in \autoref{fig:unet}(c).

The two neural network models, Unet3D$_{B}$ and Unet3D$_{Z}$, are trained together to address the azimuthal ambiguity and geometric height problems simultaneously. During each training epoch, these models generate predictions by processing the input data through their respective network architectures. The multiple loss functions defined in \autoref{eq:loss} are successively minimized to simultaneously achieve each training objective.


\begin{figure*}[t!]
\centering
\fig{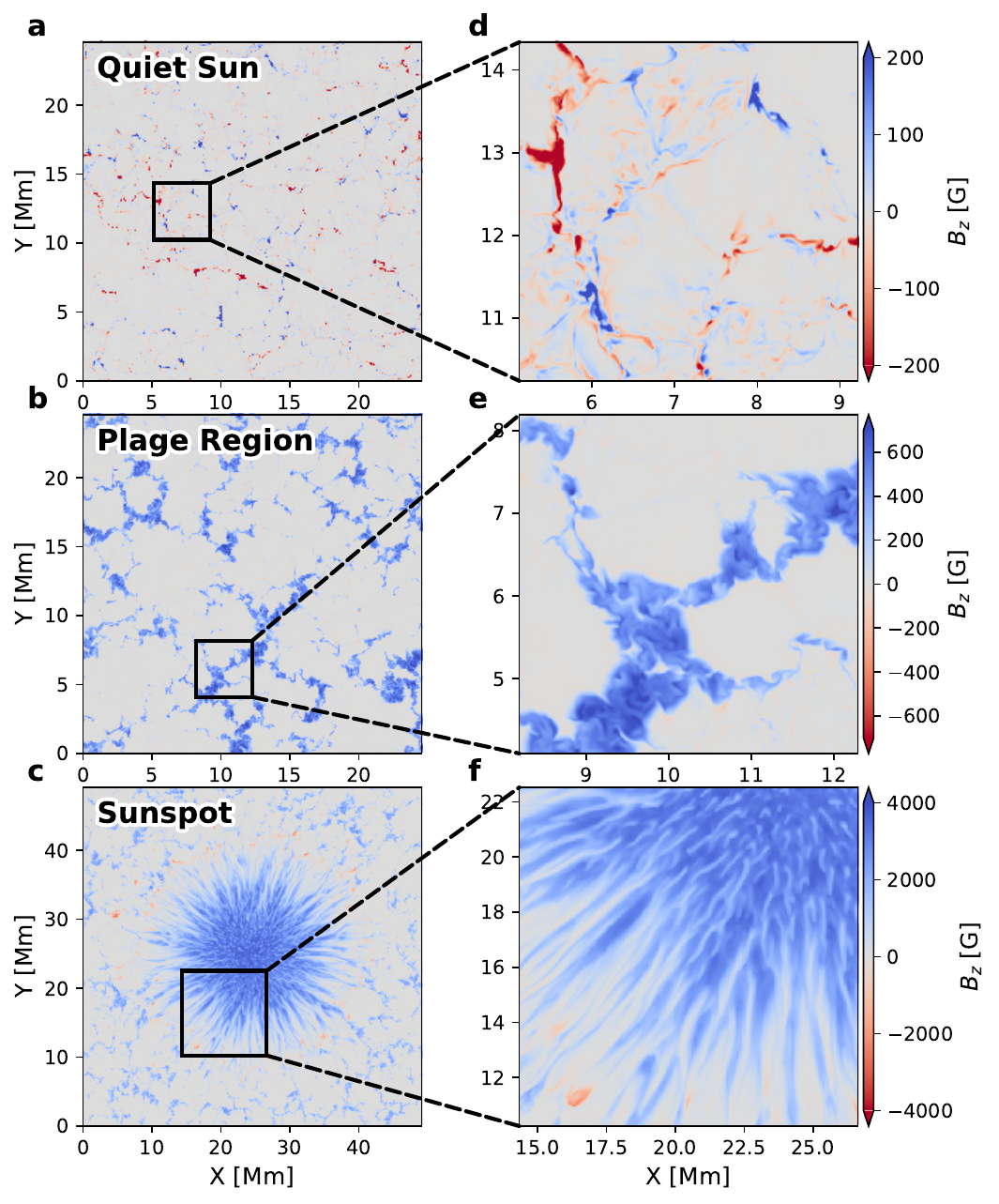}{0.6\textwidth}{}
\caption{MURaM simulations used for this study. Panels (a)--(c) present the vertical components of the magnetic field, $B_z$, on the $\log_{10}\tau=-1$ layer, for the quiet Sun, the plage region, and the sunspot runs. Panels (d)--(f) show the subregion tested in \autoref{sec:test}. \label{fig:3}}
\end{figure*}


\begin{figure*}[t!]
\centering
\fig{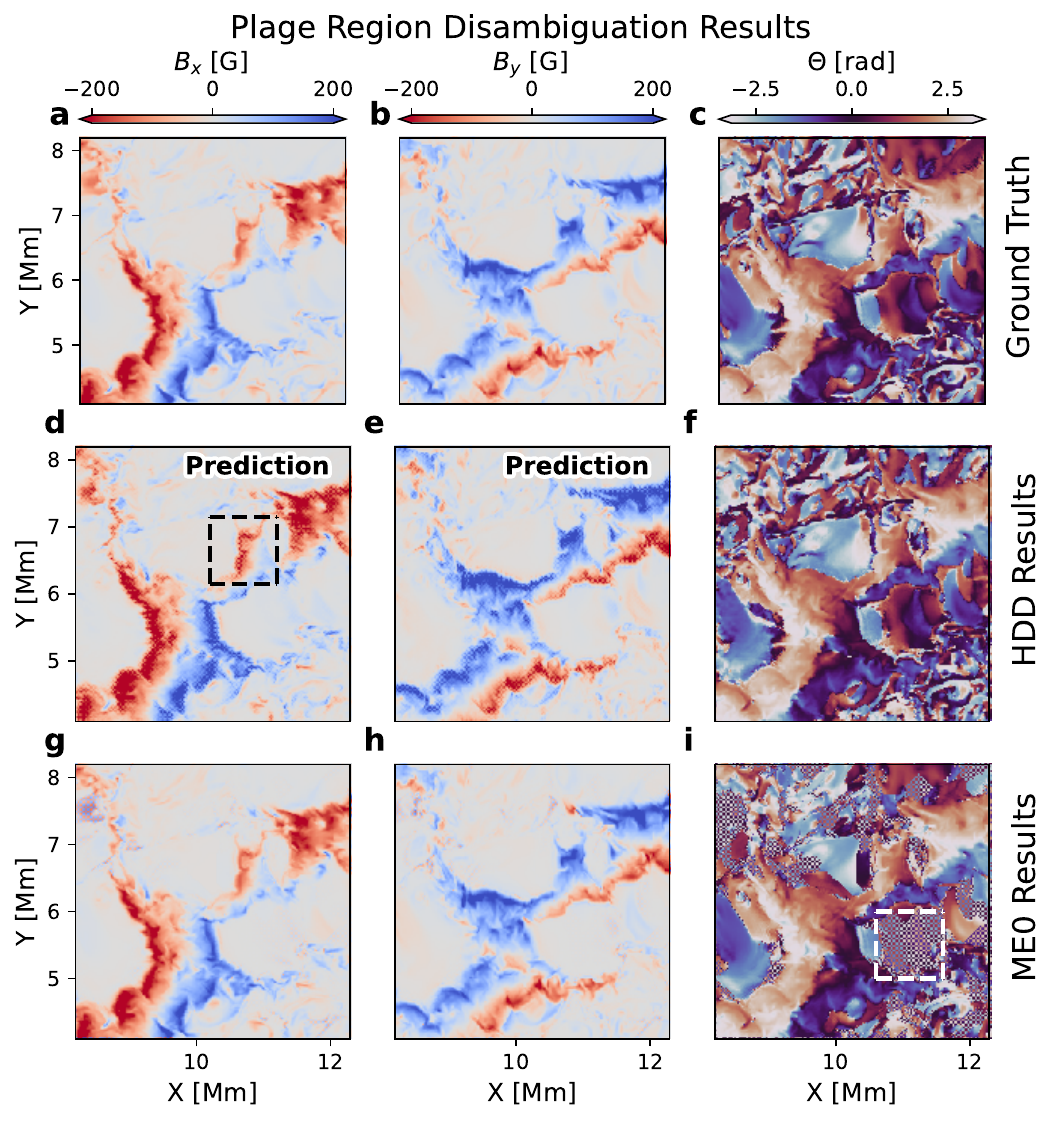}{0.62\textwidth}{}
\caption{Disambiguation results for the plage simulation on $\log_{10}\tau=-1$, the same region as \autoref{fig:3}(e). Columns, from left to right, show disambiguated $B_x$, $B_y$, and azimuth angle $\Theta$. Rows, from top to bottom, present the ground truth, HDD results, and ME0 results, respectively. HDD's $B_x$ and $B_y$ in Panels d and e are the direct outputs of the neural network, whereas $\Theta$ represents the final result after post-processing. Selected regions with ``checkerboard'' pattern are highlighted with black and white boxes in Panels d and i.\label{fig:5}}
\end{figure*}


The backpropagation \citep{1986Natur.323..533R} method is employed to compute the gradients of the total loss with respect to the model parameters. Before the optimizer updates the parameters, an adaptive gradient clipping technique is applied to improve numerical stability by preventing issues like exploding gradients. The technique maintains an exponential moving average (EMA) of gradient values, which determines the adaptive clipping thresholds based on the historical behavior of the gradients.

We use the AdamW optimizers \citep{loshchilov2018decoupled} to update the model parameters based on the computed gradients. The AdamW optimizer is an extension of the Adam optimizer \citep{KingBa15} that includes decoupled weight decay regularization, which helps prevent overfitting by penalizing large weights.

The learning rates for these optimizers are managed by the \texttt{torch.optim.lr\_scheduler.StepLR} schedulers. The learning rate is initialized at $10^{-3}$ and decays by a factor of 0.5 every 2000 epochs throughout the training process, which spans a total of 30,000 epochs. The step-wise learning rate schedule allows for larger learning rates during the initial phases of training to facilitate rapid learning and exploration of the parameter space. As the training progresses, the reduction in the learning rate enables finer adjustments to the model parameters, promoting convergence and helping to avoid overshooting minima in the loss landscape.

The aforementioned adaptive gradient clipping, dynamic learning rate scheduling, and advanced optimization techniques collectively balance stability and efficiency of the training process. The approach ensures that the models not only fit the training data, but also generalize effectively, while adhering to the physical constraints of the problem domain.

After training, to obtain the network predictions for the entire dataset, the networks are applied to the split input patches, then we assemble the output of these patches back into data cubes with the original dimensions. For each overlapping pixel, we adopt the value from the patch to which it is spatially closer. For instance, with an overlap width of 16 pixels, we used here, the first 8 pixels are taken from the first patch, and the last 8 from the next patch. This effectively eliminates an 8-pixel margin along the borders of each patch, a procedure that should suppress edge artifacts introduced by padding.


\begin{figure*}[t!]
\centering
\fig{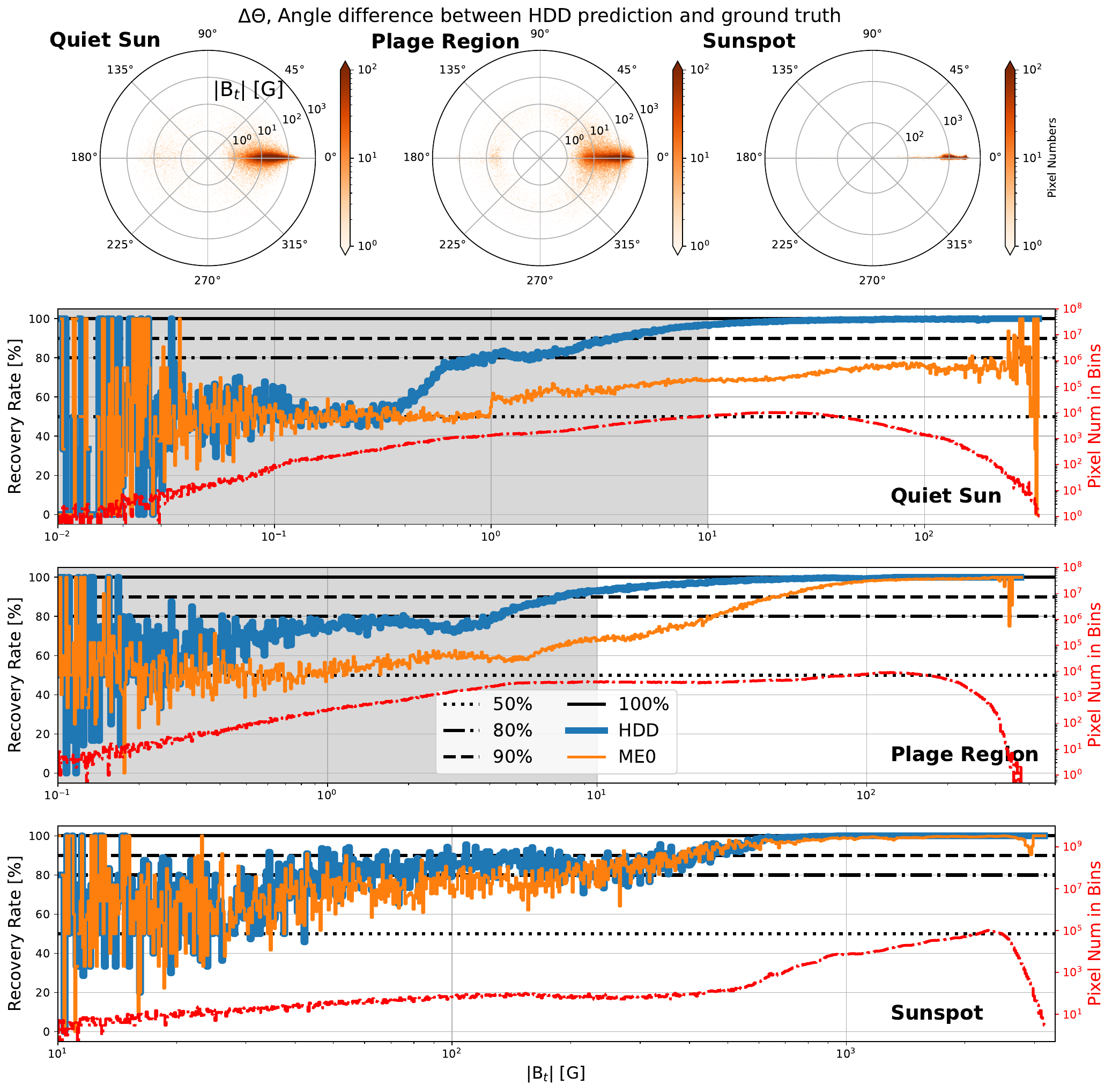}{0.75\textwidth}{}
\caption{Summary of model performance. The top row displays the 2D histogram in a polar coordinate, for the azimuth angle differences between the HDD prediction and the ground truth $\Delta\Theta = \Theta_{\rm pred} - \Theta_{\rm true}$ (as position angle), and the ground truth transverse magnetic field strength $|B_t|$ (as radius). From left to right, the plots show the results for quiet sun, plage region, and sunspot, respectively. The subsequent rows show the azimuth angle recovery rate as a function of transverse field strength for the three MHD cases, for both HDD (blue) and ME0 (orange) results. The red dash-dot line represents the pixel counts within the horizontal bins corresponding to the values of $|B_t|$. The black solid, dashed, dash-dot, and dotted lines indicate the $100$ \%, $90$ \%, $80$ \%, and $50$ \% correct recovery rate, respectively. \ky{The gray areas indicate the range where the transverse magnetic-field strength is below the estimated sensitivity of DKIST.} \label{fig:7}}
\end{figure*}


\section{Model Performance}\label{sec:test}


To assess the performance of our method, we apply it to a single snapshot each from three MURaM simulation runs, of the quiet Sun and plage regions \citep[selected from our SPIn4D dataset;][]{Yang2024spin4d}, and a sunspot \citep{Rempel2012ApJ...750...62R}. The quiet-Sun and plage simulations have a transverse pixel size ${\rm d}x=16$ km, whereas the sunspot simulation has ${\rm d}x=32$ km. All simulation domains extend from the upper convection zone to the upper photosphere. For simplicity, we place the simulation data at the solar disc center, such that the $\hat{\bm{z}}$, $\hat{\bm{x}}$, and $\hat{\bm{y}}$ directions in our analysis (i.e., the LOS and image plane axes) correspond directly to the vertical and horizontal directions in the MHD simulations.
The off-disc-center case will share the same workflow, due to the capability of the image-plane coordinate system and our self-adjust geometric height prediction, and will be tested in a future paper.

\autoref{fig:3}(a)--(c) displays the $B_z$ distribution on the $\log_{10}\tau=-1$ layer for the three cases. For each case, a small sub-region within the data is selected for detailed testing, as shown in \autoref{fig:3}(d)--(f). The 3D magnetic fields from the simulations are linearly interpolated to 32 optical depth layers between $\log_{10}\tau = 0$ to $-3.1$ with a spacing of $0.1$, and subsequently normalized to the maximum magnetic field strength within the selected region for each case. This results in a $(N_x,N_y,N_\tau) = (256,256,32)$ data cube for quiet Sun and plage regions, and a $(N_x,N_y,N_\tau) = (384,384,32)$ data cube for the sunspot case. The FOV of the selected regions are large enough to encompass multiple granules, or to include large portions of the umbra and penumbra in the sunspot. The optical depths span a range that can be (and have often been) probed by spectropolarimetric observations. The Wilson depression effect is present in both granulation and sunspot structures.

As mentioned earlier, we split each datasets into batches of $(n_x,n_y,n_\tau) = (96,96,32)$, with 16-pixel overlap between neighboring batches in both $\hat{\bm{x}}$ and $\hat{\bm{y}}$ directions (\autoref{fig:unet}(c)). To emulate the azimuthal ambiguity present in the spectropolarimetric inversion results, all azimuthal angles are restricted from the real $0$ to $2\pi$ range to the ambiguous $0$ to $\pi$ range.
Further, we supply the network with 
two versions of the ambiguity: one with the range [$-\pi/2$, $-\pi/2$] and the other with [$0$, $\pi$].  These are the first two channels of input as detailed in Section~\ref{sec:model}. 
The initial guess of $Z$ values are stored in the last channel of the input. The values are uniformly spaced with a grid size of $14.4$~km, and monotonically increase along the $z$-axis. The initial guesses of $Z$ are identical for all batches, all channels, and all $(X,Y)$ positions, as described in Section~\ref{sec:model}.


Below, we present results for the plage region as our primary example, along with brief summaries for the quiet-Sun and sunspot cases. Detailed results for the quiet Sun and sunspot are provided in \autoref{sec:a0}. Furthermore, we evaluate the generalization ability of the trained networks by applying them to unseen datasets. The results of this test are shown in \autoref{sec:a1}. To emulate more realistic conditions, we also apply our method to a rebinned dataset with added random noise. The details of the rebinning, noise injection, post-processing procedures, and HDD applications are provided in \autoref{sec:a2}. 

We additionally perform a single run with the ME0 technique \citep{Metcalf1994ApJ...428..860M} on the same datasets for comparison. As briefly mentioned in \autoref{sec:intro}, the ME0 method is widely used for resolving the 180-degree azimuth ambiguity in inversion results on a single 2D plane, typically for a single optical depth layer, or from the Milne-Eddington model \citep{Landolfi1982SoPh...78..355L} that represents an average over a range of heights \citep{WestendorpPlaza1998ApJ...494..453W}. However, the implementation of ME0 may be less optimal when applied directly to constant $\tau$ layers, because it was originally designed on a geometric constant height layer. For Milne-Eddington inversions, these constraints might already be smoothed over the height range, but for single $\tau$ layers, the $\tau-z$ mapping uncertainties can reduce its effectiveness. Here we simply run ME0 for all $N_\tau=32$ layers individually with the same settings. We note that this approach is not necessarily optimal given the input; it is intended to provide a baseline reference only. A summary of the comparison results is presented in this section. Details of the ME0 run and further discussion are presented in Section~\ref{sec:compare_me0} and \autoref{sec:me0conf}. 


\begin{figure*} [t!]
\centering
\fig{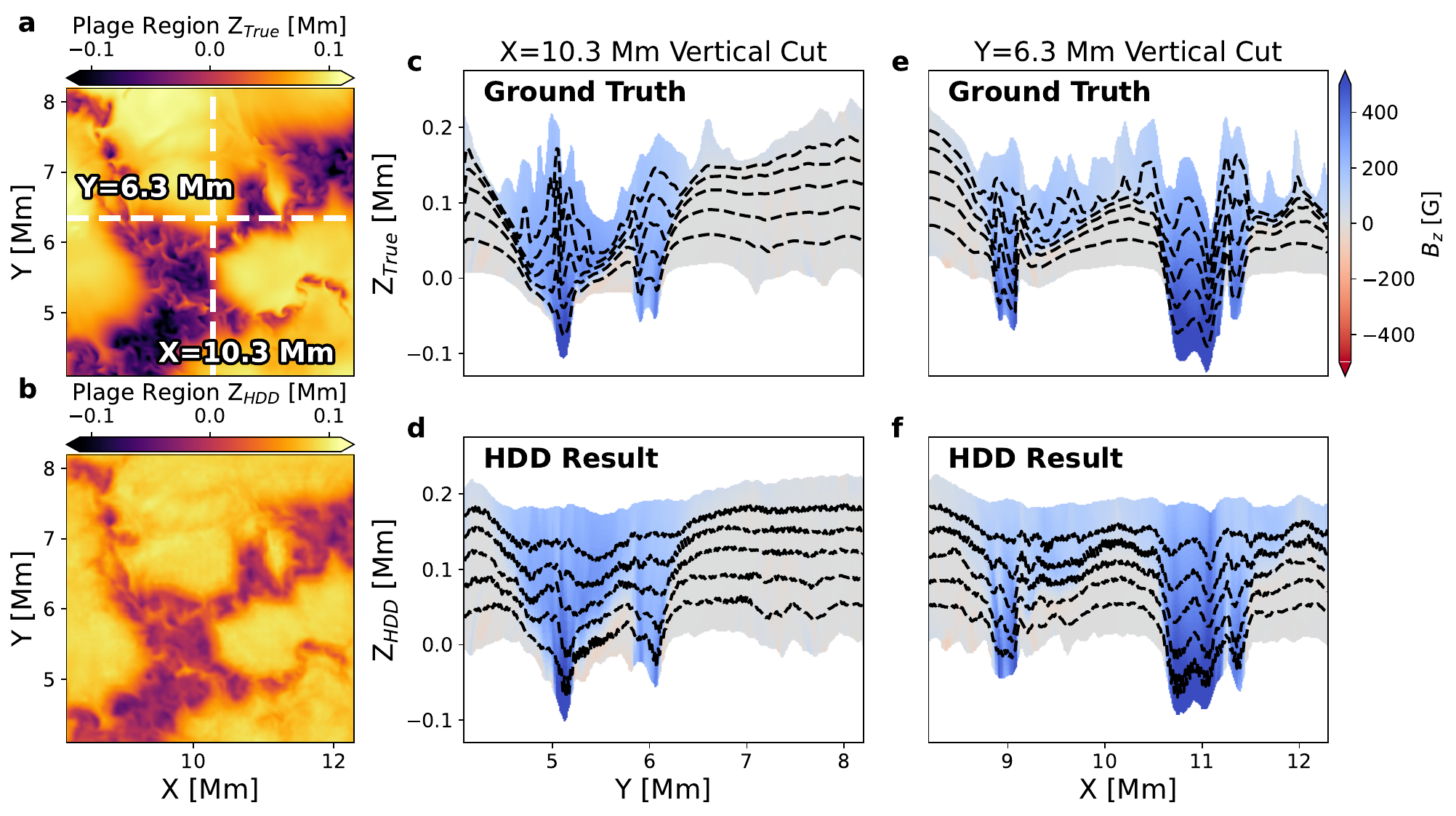}{0.8\textwidth}{}
\caption{Results for geometric heights for the plage regions. Panels (a) and (b) present the true and predicted height values for the $\log_{10}\tau=-1$ layer. Panels (c)--(f) show vertical cuts across the domain, along the dashed lines in (a). The colors represent the $B_z$ distribution and span the optical depths from $\log_{10}\tau = 0$ to $-3.1$. The five dashed black curves indicate $\log_{10}\tau$ values of $-0.5$, $-1$, $-1.5$, $-2.0$, and $-2.5$ from bottom to top, respectively. Panels (c) and (e) display ground truth at vertical cuts $X=5.6$ Mm and $Y=12.9$ Mm, respectively, while (d) and (f) show the corresponding HDD results. \label{fig:9}}
\vspace{3mm}
\end{figure*}


\begin{deluxetable}{ccccccc}	
\tablecaption{Summary of HDD and ME0 Performance\label{tab:compare}}
\tablehead{
\colhead{} & \multicolumn{2}{c}{HDD} & \colhead{} & \multicolumn{2}{c}{ME0} \\
\cline{2-3} \cline{5-6} \\[-3ex]
\colhead{Noise?} & \colhead{no\tablenotemark{\textup{*}}} & \colhead{yes\tablenotemark{\textup{\textdagger}}} & \colhead{} & \colhead{no\tablenotemark{\textup{*}}(masked\tablenotemark{\textup{\textdaggerdbl}})} & \colhead{yes\tablenotemark{\textup{\textdagger}} (masked\tablenotemark{\textup{\textdaggerdbl}})}
}
\startdata
Quiet Sun & & & & & \\
$\mathscr{M}_{\rm area}$  & 0.947  & 0.728 & & 0.688 (0.805)& 0.631 (0.748)\\
$\mathscr{M}_{J_z}$ & 0.958  & 0.737 & & 0.635 (0.805) &  0.573 (0.676)\\
$\mathscr{M}_{\Delta B}$ & 0.572  & 6.147 & & 12.28 (22.16) &  14.81 (23.21)\\
\hline
Plage & & & & & \\
$\mathscr{M}_{\rm area}$  & 0.952  & 0.846 & & 0.839 (0.774)& 0.801 (0.826)\\
$\mathscr{M}_{J_z}$ & 0.957  & 0.852 & & 0.839 (0.666) & 0.706 (0.740) \\
$\mathscr{M}_{\Delta B}$ & 0.940  & 4.013 & & 5.692 (31.17) & 11.56 (27.83)\\
\hline
Sunspot & & & & & \\
$\mathscr{M}_{\rm area}$  & 0.999  & 0.999 & & 0.993 (0.994) & 0.987 (0.988)\\
$\mathscr{M}_{J_z}$ & 0.992  & 0.998 & & 0.974 (0.982) & 0.805 (0.810) \\
$\mathscr{M}_{\Delta B}$ & 0.537 & 0.216 & & 63.32 (177.6) & 209.1 (142.3)
\enddata
\tablecomments{
\tablenotetext{\text{*}}{The test datasets without noise are presented in \autoref{sec:test}.}
\tablenotetext{\text{\textdagger}}{The test datasets with noise are detailed in \autoref{sec:a2}.}
\tablenotetext{\text{\textdaggerdbl}}{The values in the brackets are calculated within the 90\% confidence mask of ME0, as described in \autoref{sec:me0conf}.}
}
\end{deluxetable}


\subsection{Disambiguation}


The transverse magnetic field components $B_x$, $B_y$ and the azimuth angle $\Theta$ are shown in \autoref{fig:5} for the plage region case on the $\log_{10}\tau=-1$ layer as an example for the test. The transverse components in the HDD results (\autoref{fig:5}(d) and (e)) are direct outputs from the neural networks. The azimuth angle (\autoref{fig:5}(f)) is the final result from post-processing, as described in Section~\ref{sec:model}. For ME0, all three variables are taken directly from the output.

Visual inspection indicates that in most regions with relatively strong magnetic field, the azimuth angle are correct for both HDD and ME0 methods. 
Certain regions in the HDD $B_x$ and $B_y$ maps exhibit a subtle ``checkerboard'' pattern (alternating strong and weak values in space), highlighted by the black box in \autoref{fig:5}(d). Such a pattern is likely unphysical and suggests room for improvement of the neural networks. We however find that it does not significantly change the final choice of the ambiguity resolution. This is because a simultaneous change in $B_x$ and $B_y$ fortuitously leaves the predicted azimuth little affected, that is, the azimuth used for reference during post-processing (see Section~\ref{sec:model}). This surprising outcome lends empirical support to our prediction-post-processing approach over adopting the predicted magnetic field vectors directly.

The checkerboard pattern also appears in the ME0 $\Theta$ map, and can be extensive, especially in weak-field regions, highlighted by the white box in \autoref{fig:5}(i). The HDD result, on the other hand, is much smoother, and its checkerboard looks like a dithering of magnetic transverse components rather than the 180 degree rotation in ME0's azimuth angle. The pattern is a known artifact of ME0 that can be alleviated by fine-tuning the model parameters. In practice, these regions typically have small polarization signals, and are not of particular interest for the datasets to which ME0 is applied without further modification. ME0 identifies the solutions in these regions as unreliable in the ``confidence mask'' (see \autoref{sec:me0conf} for details).

We evaluate the model performance over the entire 3D volume for all three MHD cases and summarize the result in \autoref{fig:7}. The first row presents 2D histograms for the difference between the network-predicted (reference) azimuth and the ground truth, $\Delta\Theta = \Theta_{\rm pred} - \Theta_{\rm true}$, versus the true transverse magnetic field strength for each case. It is clear that, for all three cases, most pixels cluster around zero $\Delta \Theta$, indicating a close match between predicted and true angles. From the second to the last rows of \autoref{fig:7}, we assess the recovery rate of the ambiguity resolution as a function of the transverse magnetic field strength for HDD (thick blue) and ME0 (thin orange). This is calculated as the ratio between the number pixels with $|\Delta\Theta| < 90^\circ$ (such that the correct ambiguity resolution is selected during post-processing) to the total number of pixels in each bin, with the ratio given as a percentage on the left axis and the bin size indicated as red dash-dotted line, on a log scale, with reference right axis.

For the quiet Sun and plage cases, HDD method achieves near perfect results in locations with strong transverse fields, and a recovery rate exceeding $90\%$ for $|B_t|>3$ G and $|B_t|>7$ G, respectively. The ME0 recovery rates are relatively low, as expected based on the example in \autoref{fig:5}. In the sunspot case, where the magnetic field is generally stronger, the disambiguation performance is comparable between HDD and ME0 for almost all values with $|B_t|>10$~G. For reference, the transverse magnetic field sensitivity of typical DKIST observations is about $10$~G. These results tentatively suggest that the HDD can be a particularly useful tool for regions with relatively low field strengths, e.g., in the deca- to hecto-gauss range.


We quantitatively evaluate the disambiguation quality using three metrics defined in \cite{Metcalf2006SoPh..237..267M} and \cite{Leka2009SoPh..260...83L}, namely
\begin{equation}
\mathscr{M}_{\rm area}=\frac{\text{correct vertices}}{\text{total vertices}},
\end{equation}
\begin{equation}
\mathscr{M}_{J_z}=1-\frac{\sum(|J_{z,{\rm true}} - J_{z,{\rm HDD}}|)}{2\sum(|J_{z,{\rm true}}|)},
\end{equation}
\begin{equation}
\mathscr{M}_{\Delta B} = {\rm mean}(|\mathbf{B}_{\rm true} - \mathbf{B}_{\rm HDD}|).
\end{equation}
Here, $\mathscr{M}_{\rm area}$ characterizes the ratio of vertices with correct ambiguity resolution, $\mathscr{M}_{Jz}$ gives the relative absolute error in the vertical electric current density derived from the disambiguated vector field (see Section~\ref{sec:electric}), and $\mathscr{M}_{\Delta B}$ quantifies the absolute error in magnitude between the HDD solution field and the ground truth. For a perfect solution, $\mathscr{M}_{\rm area}$ and $\mathscr{M}_{Jz}$ will both be $1$, while $\mathscr{M}_{\Delta B}$ will be $0$. A summary of the comparison results across all cases is presented in \autoref{tab:compare}. Also included are results for data with added noise and for the reliable ME0 solutions identified by the confidence mask. We will discuss them in later sections.



\begin{figure*}[t!]
\centering
\fig{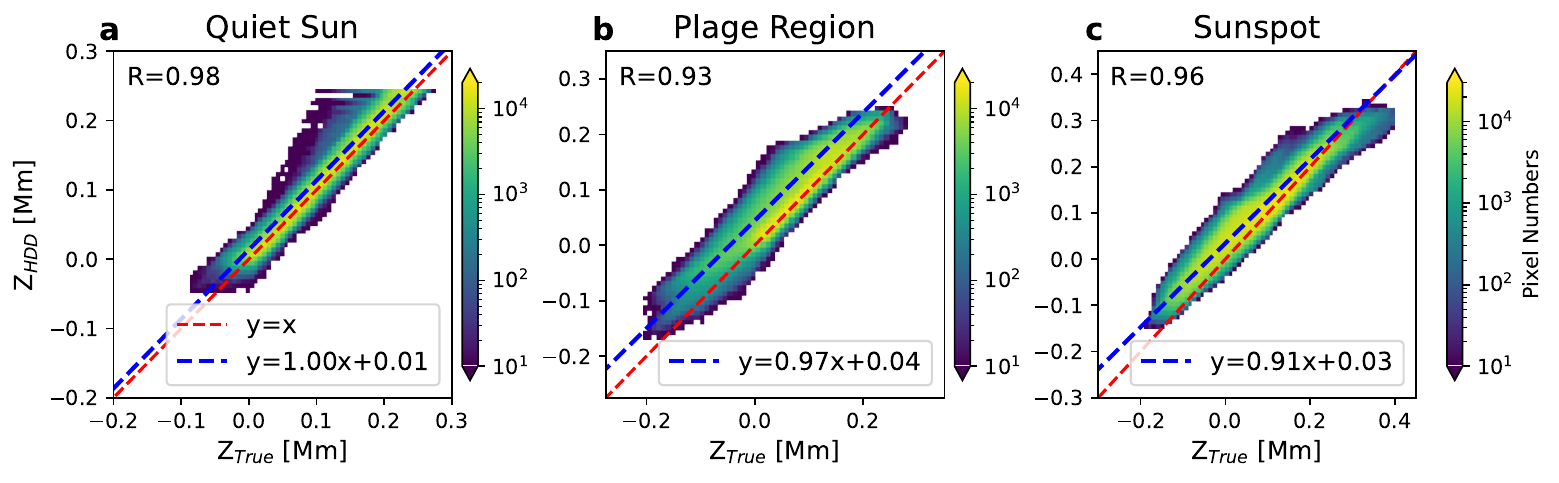}{0.8\textwidth}{}
\caption{2D histogram of predicted vs. true heights, $Z_{\rm HDD}$ and $Z_{\rm True}$, for all points across all sampled optical depths in each case. Panels (a)--(c) show results for quiet Sun, plage region, and sunspot, respectively. Red and blue dashed lines denote the identity and best linear fit (using $|B_t|$ as the weighting factor.), and $R$ marks the Pearson correlation coefficients for each case. \label{fig:11}}
\end{figure*}


\begin{figure*}[t!]
\centering
\fig{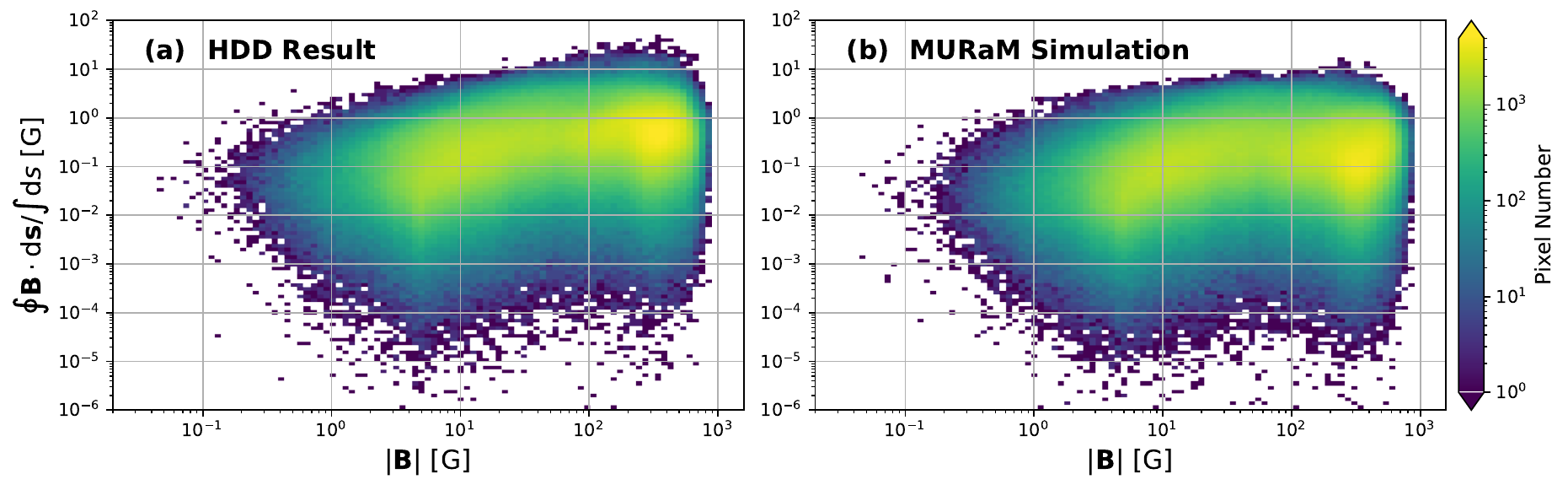}{0.75\textwidth}{}
\caption{2D histogram of the residual magnetic flux across cell surface v.s. the average magnetic field strength in the corresponding cell. (a) and (b) represent the distributions for plage from HDD's solution and MURaM simulation, respectively.}\label{fig:divb}
\end{figure*}


\subsection{Geometric Height}

\autoref{fig:9} shows an example of the predicted and true geometric heights for the plage case, including both top-down views and vertical cuts from selected slices. As the results lack an absolute zero point, we arbitrarily set the median height of the $\log_{10}\tau=0$ layer to $Z=0$ for the ground truth and the prediction, respectively.
\autoref{fig:9}(a) and (b) show the 2D distribution of geometric height for the $\log_{10}\tau = -1$. The prediction qualitatively agrees well with the ground truth. It clearly recovers the depressed heights of the intergranular lanes where the magnetic field is stronger, including various fine-scale variations therein. \autoref{fig:9} (c)--(f) further illustrate this with the height variations of the $\log_{10}\tau$ contours (lines) overplotted on $B_z$ distribution, highlighting the optical depth surface depression associated with the magnetic field. We find that the predictions for the lower layers have better agreement with the ground truth than the higher layers. We discuss the possible reason for this behavior in Section~\ref{sec:geo}. 

\autoref{fig:11} presents 2D histograms of the ground truth and the predicted heights for the entire domain, for all three MHD cases. Identity lines and $|B_t|$ weighted, least-squares linear fits are presented as red and blue dashed lines, respectively. We find that the predictions are typically within a few tens of kilometers of the ground truth, on the order of $10\%$ of the photosphere scale height. The slopes of the distribution are close to unity, specifically, $1.00$, $0.97$, and $0.91$ for the quiet Sun, plage, and sunspot, respectively. The Pearson correlation coefficients are also high, i.e., $0.98$, $0.93$, and $0.96$ for the quiet Sun, plage, and sunspot, respectively. \autoref{tab:hdd_performance} summarizes the correlation coefficients for geometric heights and several other comparisons, including the results with added noise (\autoref{sec:a2}), and those for the derived electric current density (Section~\ref{sec:electric}) and the Lorentz force (Section~\ref{sec:lorentz}).

\begin{deluxetable}{cccccccccc}	
\tablecaption{HDD Performance of 3D Variables\label{tab:hdd_performance}}
\tablehead{
\colhead{} & \multicolumn{2}{c}{$R(Z)$} & \colhead{} & \multicolumn{2}{c}{$R(J)$} & \colhead{} & \multicolumn{2}{c}{$R(F)$}\\
\cline{2-3} \cline{5-6} \cline{8-9}\\[-3ex]
\colhead{Noise?} & \colhead{no} & \colhead{yes} & \colhead{} & \colhead{no} & \colhead{yes} & \colhead{} & \colhead{no} & \colhead{yes}}
\startdata
Quiet Sun & 0.98 & 0.98 & & 0.98 & 0.80 & & 0.99 & 0.85\\
Plage & 0.93 & 0.88 & & 0.96 & 0.86 & & 0.99 & 0.94\\
Sunspot& 0.96 & 0.91 & & 0.97 & 0.91 & & 0.96 & 0.82 
\enddata
\tablecomments{
This table summarizes the Pearson correlation coefficient $R$ between the predicted and the ground truth values, for the geometric height $Z$, the magnitude of vector electric current density $J=|\bm{J}|$, and the magnitude of Lorentz force $F=|\bm{F}|$.
}
\end{deluxetable}


\begin{figure*}[!t]
\centering
\fig{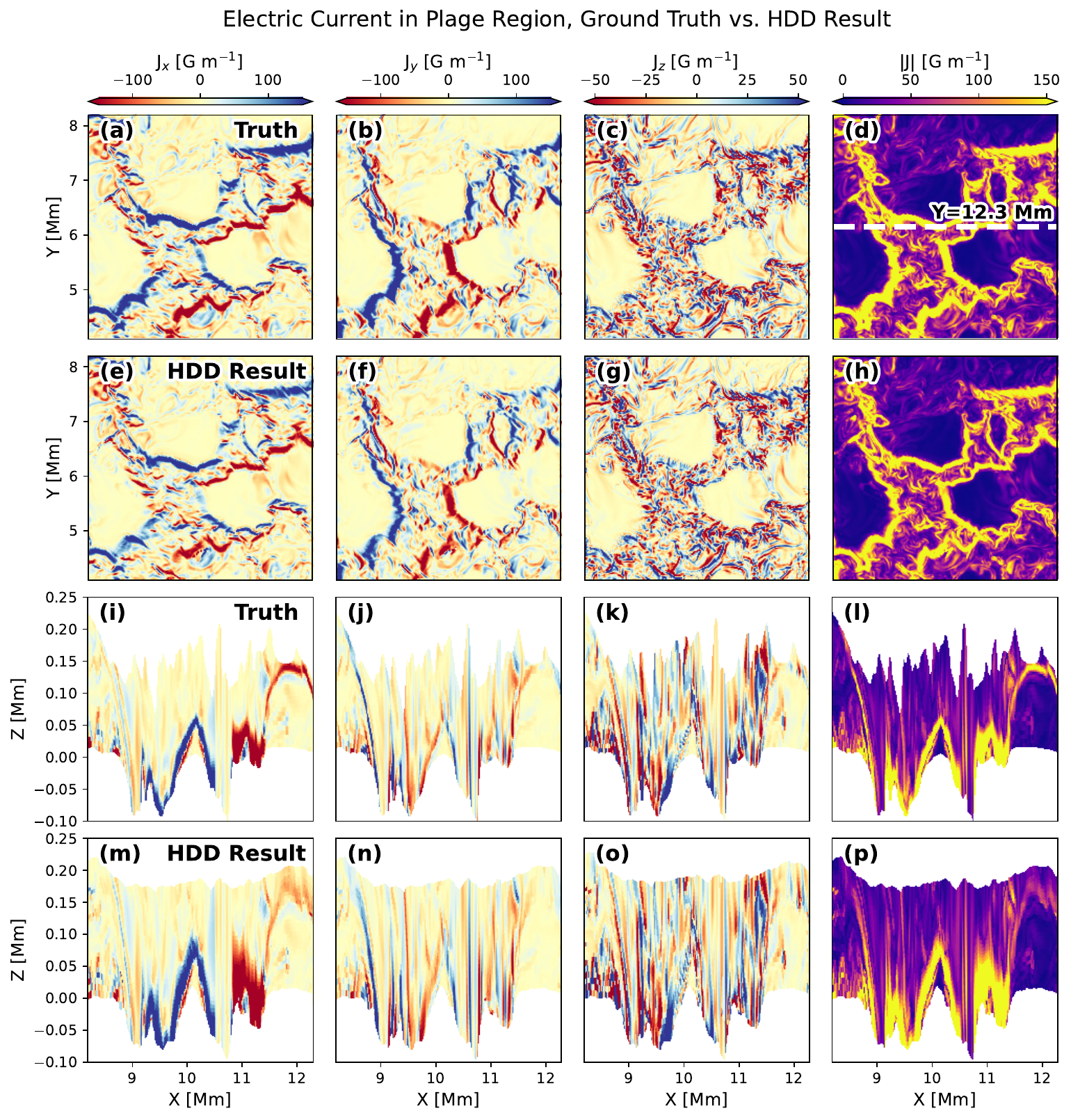}{0.75\textwidth}{}
\caption{Columns from left to right show the $x$-, $y$-, $z$-components, and the absolute value of the electric current in the plage case. Panels (a)--(d) and (e)--(h) display the top view of the electric current on the $\log_{10}\tau = -1$ layer of the ground truth and HDD result, respectively. Panels (i)--(l) and (m)--(p) present the electric current distribution on the true and predicted vertical cuts at $Y = 12.3 $~Mm, indicated by the white dashed line in panel (d). \label{fig:13}}
\end{figure*}


\subsection{Divergence}

We examine the divergence of the modeled magnetic field in the volume. This quantity is explicitly minimized during training: the residual thus characterizes how well the imposed physical constraint is satisfied. It provides an empirical test on both the azimuth ambiguity resolution and the predicted geometric height.

To this end, we calculate the residual magnetic flux across all cell surfaces for our MURaM ground truth without noise, as defined in \autoref{eq:flux}. This flux is then normalized by the total cell surface area, yielding the final metric, $\oint \bm{B} \cdot {\rm d}\bm{s} /\int {\rm d}s$. For a fair comparison, we calculate this metric in the same fashion for our result and the MURaM input (interpolated to constant $\tau$ surfaces and integrated over cell surfaces). Here and after, we opt to focus on the HDD final result after post-processing rather than the intermediate, network predicted field. A 2D histogram of this metric versus the average magnetic field strength $|\bm{B}|$ within each corresponding cell is shown in \autoref{fig:divb} for the plage case. The two distributions are visually very similar.
Across the full dataset, the median value of this magnetic charge is $0.25$~G for the HDD solution and $0.12$~G for the MURaM simulation, well below the median MURaM field strength or the typical uncertainty of real observations.
$\nabla\cdot\vec{B}$ in MURaM is appropriately calculated using a 4th order finite difference stencil at MURaM cell centers (our vertices) and has a mean/std of 0.012/0.026~G in the plage simulation over the analyzed volume. As is typical of numerical simulations, using a different stencil to calculate a numerical value of the solenoidal condition will result in substantial errors, typically 3-4 orders of magnitude larger than the code-consistent calculation. Our (inconsistent with MURaM) finite volume calculation of divB produces divergence values roughly within a factor of two of the same, numerically-inconsistent calculation using the Muram data itself.  This difference is insignificant compared to the difference between the correct and incorrect stencils applied only to MURaM data. That is, our divergence results are essentially consistent with the ground truth code, to within the numerical accuracy allowed after interpolation and applying a different numerical stencil for the calculation.





\begin{figure*}
\centering
\fig{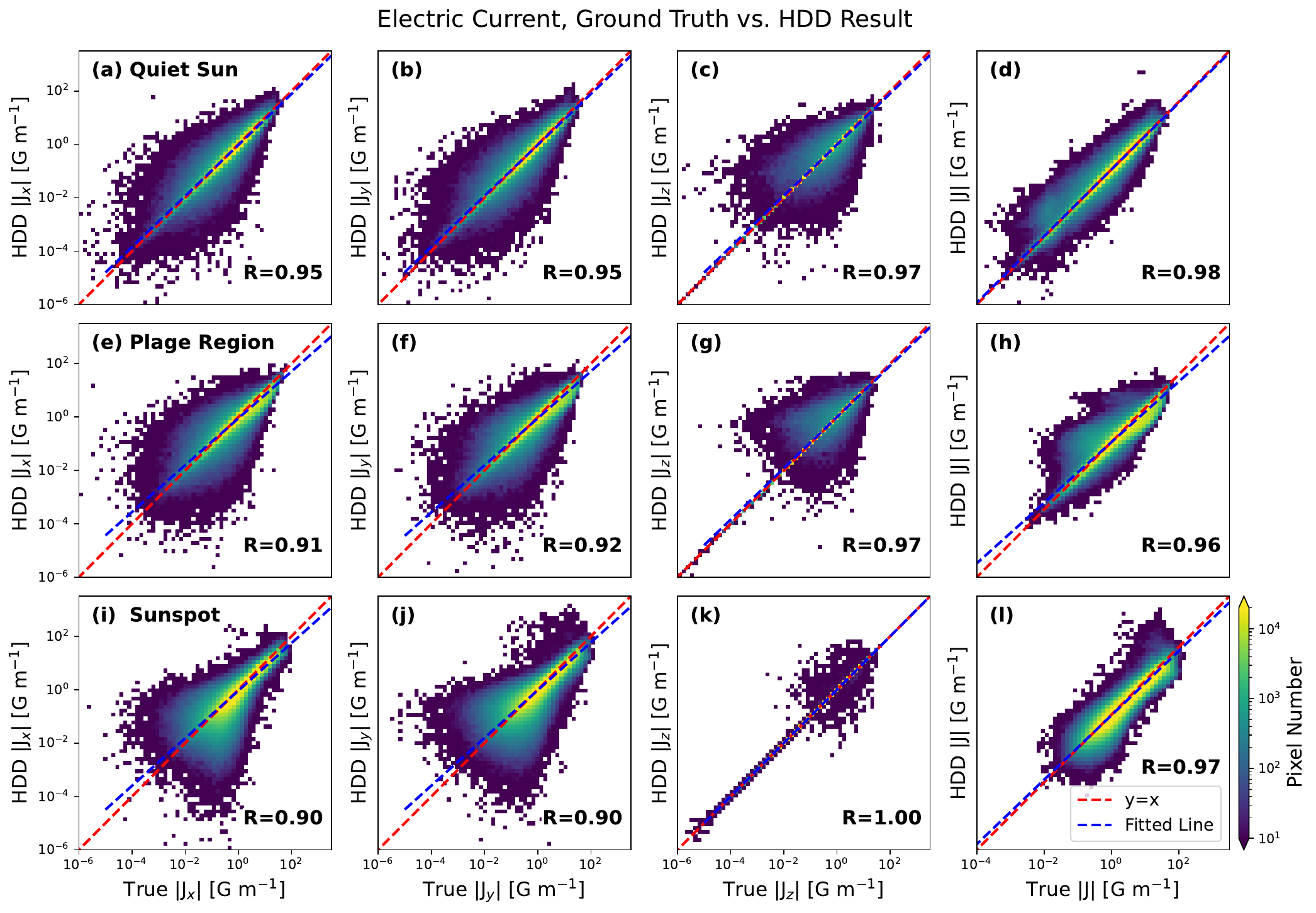}{0.78\textwidth}{}
\caption{2D histogram comparing the ground truth and the HDD output electric current density. Columns from left to right show the $x$-, $y$-, $z$-components, and absolute current magnitude, while rows from top to bottom represent results for the quiet Sun, plage region, and sunspot, with Pearson correlation coefficients specified for each case. Red and blue dashed lines denote the identity and the best-fit linear relation, respectively. \label{fig:15}}
\end{figure*}


\section{Scientific Applications}\label{sec:application}


Resolving the azimuthal ambiguity and the geometric height problems effectively yields a 3D grid of vector magnetic field with an absolute physical scale. Unlike the output from conventional inversion on a $\tau$ grid, our results directly enable the estimate of auxiliary quantities, for example, the electric current density $\bm{J}=\nabla \times \bm{B}$ (neglecting the constant $c/4\pi$) and the Lorentz force $\bm{F} = \bm{J} \times \bm{B}$, that would otherwise be inaccessible. These two quantities are of broad scientific interest. Below, we demonstrate the efficacy of the HDD method in estimating $\bm{J}$ and $\bm{F}$ for the plage case. Results for the quiet Sun and the sunspot cases are presented in the \autoref{sec:a0}.

We note that the $\bm{J}$ and $\bm{F}$ vectors discussed in this section are expressed in our Cartesian coordinate. As mentioned in Section~\ref{sec:coordinates}, our ``vertical'' direction $\hat{\bm{z}}$ typically differs from the direction of solar gravity $\bm{g}$. Coordinate transformation will be required (1) to project the vectors, and (2) to register the location of the vertices into the Heliographic frame. Below, we assume that our region of interest is at the disc center for simplicity.



\begin{figure*}[t!]
\centering
\fig{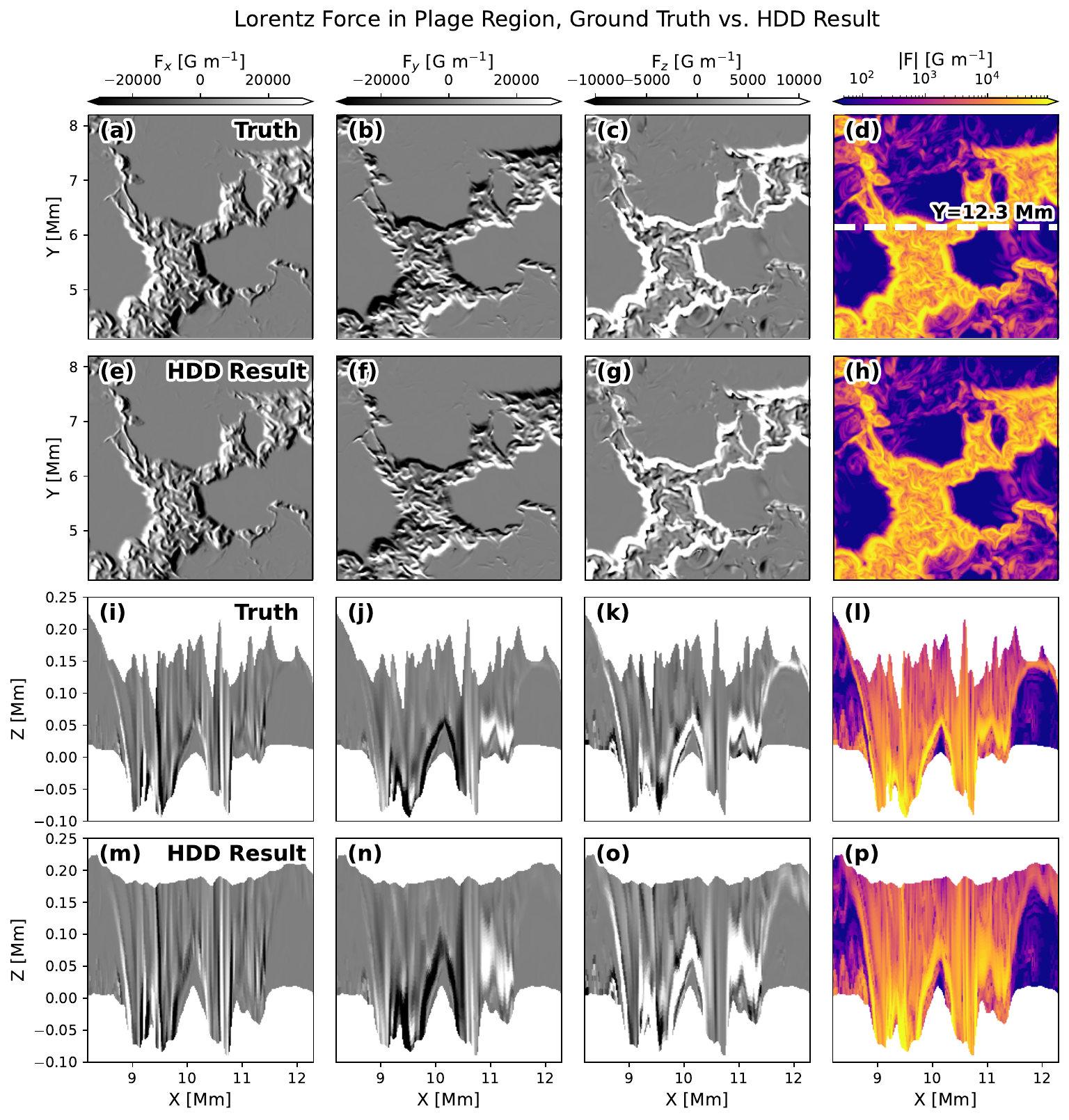}{0.75\textwidth}{}
\caption{Same as \autoref{fig:13}, but for the Lorentz force distributions. \label{fig:18}}
\end{figure*}

\begin{figure*}
\centering
\fig{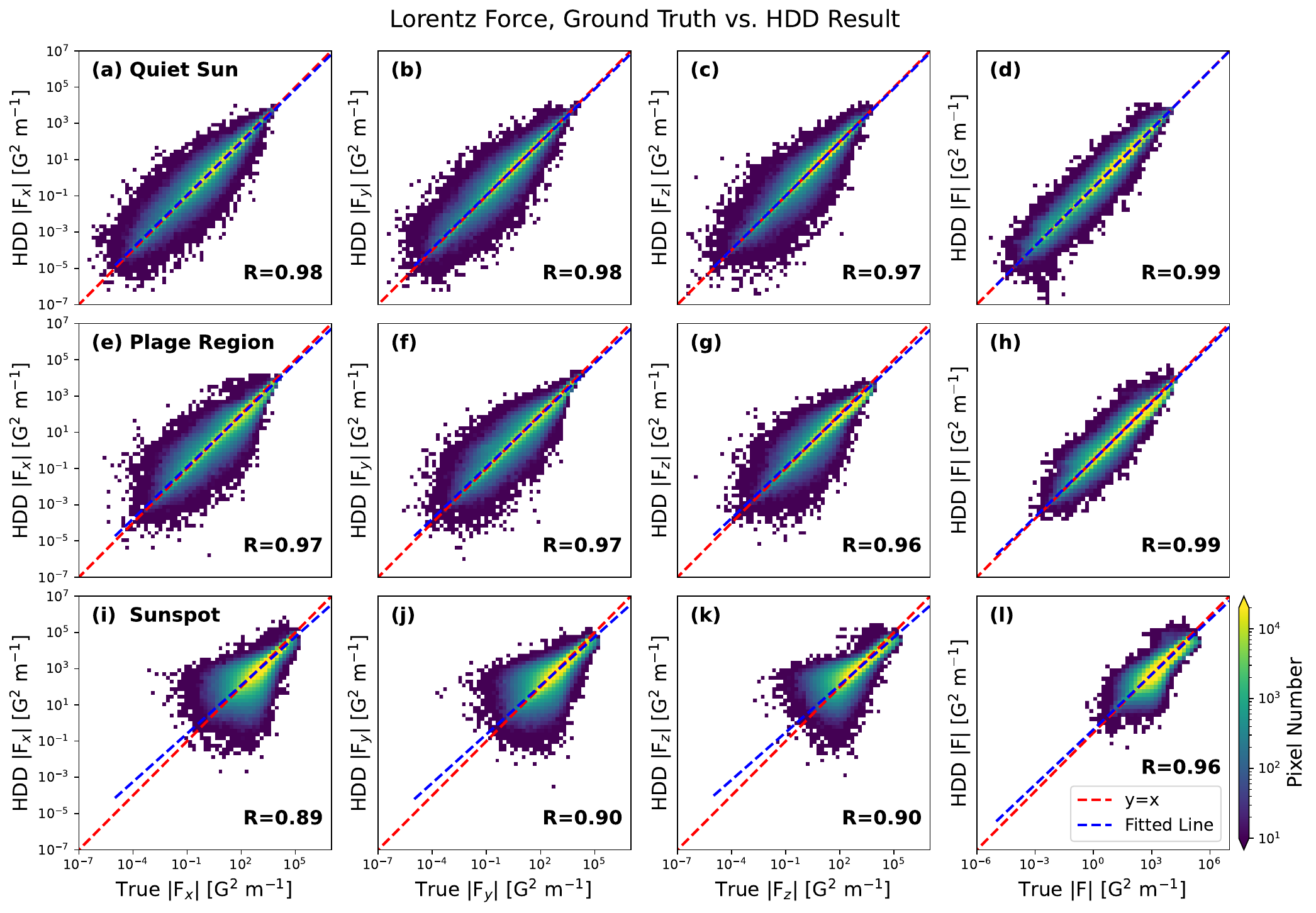}{0.78\textwidth}{}
\caption{Same as \autoref{fig:15}, but for the Lorentz force comparison. \label{fig:20}}
\end{figure*}


\subsection{Vector Electric Current in 3D Atmosphere}\label{sec:electric}

\autoref{fig:13} presents a comparison of the predicted and true vector electric current density for the plage case. Similar to the results for geometric height (\autoref{fig:9}), most interesting structures reside in the intergranular lanes, including enhanced $J_x$, $J_y$, and $|J|$ along the lane boundaries (Panels (a), (b), (d)), enhanced $J_x$ and $|J|$ at the lowest $\tau$ layers (Panel (i), (l)), and numerous opposite-polarity, sheet-like regions of $J_z$ (Panels (c), (k)). The HDD results qualitatively reproduce all them well. \autoref{fig:12} and \autoref{fig:14} present the results for the quiet Sun and the sunspot cases, from which we draw similar conclusions.

\autoref{fig:15} provides 2D histograms between the ground-truth and the HDD output electric current densities, displayed on a $\log$ color scale. All correlations are at least $0.9$, with most pixels concentrated along the identity line. We note that the correlation coefficients are for all pixels, whose values span about 7 orders of magnitude. For pixels with stronger magnetic fields that can be meaningfully inferred from observation, the correlation is even higher. \autoref{tab:hdd_performance} summarizes these Pearson correlation coefficients.

Electric currents are essential for understanding magnetic energy dissipation, which powers many dynamic processes in the solar atmosphere \citep{daSilvaSantos2022A&A...661A..59D}.
Our evaluation shows that the HDD method can accurately capture both the strength (\autoref{fig:15}) and the 3D distributions (Figures \ref{fig:13}, \ref{fig:12}, and \ref{fig:14}) of the electric current density, for three very different magnetic environments on the Sun. Our test on a dataset with reduced resolution and added noise in \autoref{sec:a2} yield similarly satisfactory results (\autoref{fig:a4}). Notably, the accuracy is highest in lower optical depth layers where the magnetic field strength is greater, similar to the geometric height prediction mentioned in Section~\ref{sec:geo}. Overall, these results demonstrate the model's capability to reconstruct 3D vector electric current structures in the lower solar atmosphere.

Other advanced methods for reconstructing the 3D atmosphere, such as FIRTEZ developed by \ky{\cite{Yabar2019A&A...629A..24P}} and \cite{Borrero2023A&A...669A.122B}, incorporate MHS assumptions (balancing Lorentz force, pressure gradient, and gravitational force) to estimate the current densities. Their results, similar to ours, are worse in higher layers. It is worth noting that a direct comparison between HDD and FIRTEZ is not entirely appropriate. This is because FIRTEZ is essentially a spectropolarimetric inversion algorithm, which introduces additional constraints in the magnetic parameters.

The 3D vector electric current density from our new technique can provide valuable insights into magnetic structures and magnetically powered dynamics under a variety of solar conditions \citep{Metcalf1994ApJ...428..860M}. For active regions, the currents inside the magnetic flux ropes are crucial to their eruptive potential \citep{Torok2014ApJ...782L..10T,Liu2017ApJ...846L...6L,Avallone2020ApJ...893..123A}, but the origin is hotly debated due to their ambiguous distribution \citep[e.g.,][]{Sun2021SoPh..296....7S}. The response of currents to flaring activities has also been investigated \citep{Janvier2016A&A...591A.141J}. However, most studies are still limited to only the vertical component $J_z$, derived from an assumed single-height magnetic field map, where only the horizontal derivatives of the field are accessible. It is unclear in which areas of the solar atmosphere the vertical currents dominate the horizontal, or if any such regions exist in general. Instead, our new method allows calculation of the full current vector, which is essential to understanding regions with strong horizontal currents. Observations suggest that horizontal currents are likely at least as strong as vertical currents in sunspots \citep{Puschmann2010ApJ...721L..58P,Nechaeva2021GeAe..61..956N}.

A unique strength of HDD is its capability to recover sheet-like current structures in our test dataset (Figures \ref{fig:13}, \ref{fig:12}, \ref{fig:14}). These sheet-like current structures are thought to be necessary for magnetic energy dissipation that heats the corona and chromosphere \citep{Priest2021A&A...647A..31P,daSilvaSantos2022A&A...661A..59D,Pontin2022LRSP...19....1P}. Past studies show that filamentary currents at chromosphere heights can form vertical sheets near magnetic boundaries such as the umbra-penumbra interface \citep{Tan2007AdSpR..39.1826T, Anan2021ApJ...921...39A}. Nevertheless, their existence and properties in the real 3D atmosphere are still under-explored. Our HDD model, coupled with high-resolution observations, can change this situation.


\subsection{Lorentz Force in 3D Atmosphere}\label{sec:lorentz}

\autoref{fig:18} presents a comparison of the predicted and true vector Lorentz force for the plage case, similar to the electric currents. Interesting structures include overall stronger $|F|$ in the intergranular lanes, and more enhanced force along the lane boundaries (Panels (a)--(d)). These are again recovered by the HDD results. \autoref{fig:17} and \autoref{fig:19} present the results for the quiet Sun and the sunspot cases, from which we draw similar conclusions.

\autoref{fig:20} provides 2D histograms between the ground-truth and the HDD output Lorentz force. Good agreement is present for values ranging over eight orders of magnitude in plage region and quiet Sun, and about four orders for the sunspot case. The Pearson correlation coefficients of the logarithmic absolute values are at least $0.96$ in plages and quiet Sun. The correlation for the sunspot case is slightly lower, with a minimum of $0.89$. \autoref{tab:hdd_performance} summarizes these Pearson correlation coefficients.

The high correlation coefficients demonstrate that our method provides a direct way to measure the 3D distribution of the vector Lorentz force quantitatively. One immediate application is to assess the degree of force-freeness of the lower solar atmosphere. It has long been known that the photosphere cannot be approximated as a force-free environment, which calls into question the validity of many force-free extrapolation methods that use photospheric magnetic field as the bottom boundary \citep[][and references therein]{Wiegelmann2021LRSP...18....1W}. Minor departures from force-freeness can result in substantially different field properties, from total energy to field-line connectivity \cite{Peter2015A&A...584A..68P}. This will also affect many analyses that depend on the force-free assumption, like the twist distribution estimated as $\alpha=J_z/B_z$ \citep{Leka1999SoPh..188....3L,Wheatland2000ApJ...532..616W,Liu2014ApJ...785...13L}, and the energy estimation based on MHD virial theorem \citep{Low1982SoPh...77...43L,Wheatland2006ApJ...636.1151W}. The chromosphere may be more force-free, but the relatively few studies so far yield contradictory results \citep{Metcalf1995ApJ...439..474M,Socas-Navarro2005ApJ...631L.167S}. Conventional methods to check the departure of the force-freeness condition usually evoke the integrated Lorentz force from a surface integral of the Maxwell stress tensor \citep{Low1985svmf.nasa...49L,Zhang2023ApJ...956L..17Z} or from a 3D reconstructed atmosphere based on a rough geometric height estimation \citep{Socas-Navarro2005ApJ...631L.167S}. In reality, the Lorentz force distribution is spatially non-uniform (Figures~\ref{fig:18}, \ref{fig:17}, and \ref{fig:19}). The HDD method, combined with high resolution and multi-line observations, can directly tackle this question in a spatially resolved fashion.

The HDD method will also allow us to study the rapid temporal evolution of the Lorentz force with high-cadence observations. A large force impulse is thought to impart magnetic imprints on the lower atmosphere during solar eruptions \citep{Wang2002ApJ...576..497W,Liu2012ApJ...745L...4L,Liu2016NatCo...713104L,Liu2018ApJ...869...21L,Sun2017ApJ...839...67S,Xu2018NatCo...9...46X}. These imprints may be attributed to the downward counterpart of the coronal Lorentz force that accelerates the eruption, acting to conserve momentum of the system \citep{Hudson2008ASPC..383..221H,Fisher2012SoPh..277...59F}, or a large-amplitude shear Alfv\'en wave propagating down from the corona \citep{Wheatland2018ApJ...864..159W}. 


\section{Discussion}\label{sec:discussion}

\subsection{Model and Training}\label{sec:dis_model_train}

In the pioneering work of \cite{AsensioRamos2019A&A...626A.102A}, neural networks (SICON) were trained on Stokes profiles generated from MURaM simulations. Applying such methods to real observations \citep{EstebanPozuelo2024A&A...689A.255E,Kriginsky2024arXiv241105532K} would require a large training dataset to encompass the vast parameter space. This challenge is also a key focus of the SPIn4D project, as outlined by \cite{Yang2024spin4d}. However, a supervised learning model trained on MURaM inherently produces predictions resembling MURaM simulations, limiting its adaptability to real observations. The problem is known as the out-of-distribution problem in supervised machine learning. 

In contrast, the HDD approach provides a framework for reconstructing the 3D lower atmosphere independent of training data. The method is based on the UNet3D architecture, which can efficiently handle 3D data structures from spectropolarimetric inversions (two spatial dimensions along the FOV and one in optical depth, the latter being isomorphic to the third, spatial dimension). By employing a learnable model, we leverage the capability of deep learning networks to capture complex spatial patterns and correlations within the data. This enables the model to flexibly adapt to various solar magnetic field configurations. This unsupervised learning approach essentially operates as a learnable PDE solver.

However, this flexibility also introduces the disadvantage of multiple hyperparameters within the custom loss functions (\autoref{eq:loss}). The performance of the UNet3D architecture is sensitive to the hyperparameter settings such as the weights (\autoref{tab:weight}), learning rates, regularization coefficients, batch sizes, and network architecture parameters (e.g., number of layers and channels). Optimizing these hyperparameters requires careful tuning and can be computationally intensive due to the need to balance various, heterogeneous loss function terms. As an initial attempt, we select the parameters in this work via extensive testing by trial and error. This falls short of the typical hyperparameter optimization in machine learning, which requires a systematic parameter survey. Future research could focus on systematically exploring the hyperparameter space using techniques like grid search or Bayesian optimization to achieve optimal results.

For this work, we primarily use the \texttt{ResidualUNet3D} module. As we have not sufficiently tested it with different layers, channels, activation functions, and normalization layers, we cannot rule out the possibility that other architectures may provide better performance.

The training requires tens of thousands of epochs, significantly more than conventional, supervised UNet training. For our tests in \autoref{sec:test}, each case took approximately 750~minutes using two V100 GPUs, compared to around 250 minutes per case with ME0 on Intel(R) Xeon(R) CPU E5-2680. Nevertheless, as the HDD method is based on PyTorch with a fully parallelizable design, leveraging more GPUs and advanced acceleration strategies can significantly speed up the training.


\subsection{Disambiguation}\label{sec:disamb}

In sunspots, the transverse field strength varies significantly (e.g., along the red dash-dotted lines in \autoref{fig:7}). Values can differ by about three orders of magnitude, and there are much fewer weak-field pixels. This results in an imbalanced training sample, and therefore larger fluctuations in the recovery rate compared to quiet-Sun and plage regions in the weak-field regions.

The divergence in our model is approximated by the net residual magnetic flux across all the surfaces for each mesh cell (\autoref{fig:1}(a)). Our flexible, deformable coordinate system allows for an easy implementation of the flux calculation. This approach overcomes the challenges of computing numerical derivatives on a corrugated plane, as expected due to the Wilson depression. The accuracy of the approximation is expected to improve with smaller grid size. As such, this model is well-suited to take advantage of the capabilities of large ground-based telescopes like DKIST, Goode Solar Telescope \citep{Cao2010AN....331..636C}, the future European Solar Telescope \citep[EST,][]{QuinteroNoda2022A&A...666A..21Q}, and Chinese 2.5 m wide-field and high-resolution telescope \citep[WeHoST,][]{Fang2019SSPMA..49e9603F}.

Most disambiguation algorithms search for a binary solution, i.e., whether $180^\circ$ should be added to the azimuth. The discrete nature of this formulation does not work well with machine learning and can easily cause the training to fail. Instead, we rewrite the task in the form of minimizing continuous, differentiable functions such as ${\rm loss}_{B_t}$ and ${\rm loss}_{\rm parallel}$ that produce a continuous predicted azimuth, and then use $|\Delta\Theta|>90^\circ$ as a binary discriminator to resolve the ambiguity. If the input magnetic field to the HDD were exactly correct \emph{except} for the $180^\circ$ ambiguity, a perfect solution would maintain the magnitude of the transverse field, and produce an azimuth either parallel or antiparallel to the input. Small, numerical deviations in the predicted azimuth would not affect the choice of the binary solution in the post-processing step, except very close to $\Delta\Theta=\pm 90^\circ$, which should never arise. In practice, we indeed find that vanishingly few locations suffer from that problem (see histograms in \autoref{fig:7} top row). This strategy leads to a stable algorithm that maintains the best-fit (but ambiguous) solution for an input spectropolarimetric inversion, and proves to work well for our purpose.

The unsupervised nature of HDD enables flexible applications. Examples include: (1) it can be used directly on data without pre-training; (2) it can be pre-trained on simulations or observations, and then fine-tuned for new observations; (3) it can be pre-trained on simulations and applied directly to observational data. An example is presented in \autoref{sec:a1}, where the model pre-trained in \autoref{sec:test} are directly applied to data outside the FOV of the test datasets. The method also performs well with small datasets, such as the $16$ or $36$ batches used in our tests. 


We note that the solution space of the ambiguity problem is extremely large, even with only two possible orientations per pixel. For a grid with approximately $100 \times 100 \times 30$ pixels in our test cases, this results in $2^{3 \times 10^5} \approx 10^{10^5}$ possibilities. For real observations of $1000 \times 1000 \times 30$, the value increases to about $10^{10^7}$. For the latter case, we split the input data into many patches, which significantly reduce the overall solution space. 

The performance of the HDD method will benefit from high-quality spectropolarimetry observation and accurate inversion. This is because the reconstruction relies heavily on the data inherently satisfying the divergence-free condition. The information of the Wilson depression is encoded in the vector magnetic field and its relation with the neighboring pixels; a more accurate inversion is expected to produce a better final solution. 

To assess the robustness of our method, we conduct a test using a noisy, low-resolution dataset, as discussed in \autoref{sec:a2}. The performance somewhat deteriorates as expected, but the reconstructed 3D electric currents and Lorentz forces are still of reasonable quality where $B_t$ exceeds the approximated DKIST sensitivity (shown in \autoref{fig:7} and \autoref{fig:a2}). \ky{We note, however, that the spatial-resolution reduction adopted in \autoref{sec:a2} is rather simplistic. In reality, the effect is characterized by point-spread function (PSFs) that include contributions from both the telescope optics and atmospheric turbulence. They will attenuate polarimetric signals and mix light over distances. The impact is expected to be severe where the magnetic topology varies much on small spatial scales. In such cases, the divergence-free assumptions of the magnetic field may be intrinsically violated in the inversion result. Future work should treat the PSF explicitly at the inversion processing, either via deconvolution when a reliable PSF is available \citep{DiazBaso2018AA...614A...5D,Korda2025AA...697A..28K}, or, preferably, by incorporating a spatially coupled forward model that convolves synthetic Stokes spectra with the PSF during inversion \citep{vanNoort2012AA...548A...5V,CastellanosDuran2024A&A...687A.218C}.} 





\begin{deluxetable*}{llcl}
\tablecaption{ME0 parameters\label{tab:me0}}
\tablewidth{0pt}
\tablehead{
  \colhead{Parameter} & \colhead{Value} & \colhead{Parameter} & \colhead{Value}
}
\startdata
irflag   & 0                    & nerode   & 0 \\
ibflag   & 0                    & ngrow    & 0 \\
iaflag   & 0\tablenotemark{a}   & iaunit   & 1 \\
igflag   & 1\tablenotemark{b}   & ipunit   & 0 \\
ipflag   & 0                    & incflag  & 0 \\
npad     & 200\tablenotemark{c} & iseed    & 2 \\
nap      & 10                   & iverb    & 1 \\
ntx      & 10                   & neq      & 100 \\
nty      & 10                   & $\lambda$& 1 \\
athresh  & 1, 10\tablenotemark{d} & tfac0  & 2 \\
bthresh  & 1, 10\tablenotemark{d} & tfactr & 0.99\tablenotemark{e} \\
\enddata
\tablecomments{Parameters used in the ME0 configuration.}
\tablenotetext{a}{Zero-azimuth direction is along $+\hat{\boldsymbol{x}}$.}
\tablenotetext{b}{Computation uses planar geometry.}
\tablenotetext{c}{Number of pixels added to each side of the field of view.}
\tablenotetext{d}{Thresholds to activate simulated annealing: $1$\,G for quiet Sun/plage and $10$\,G for sunspots.}
\tablenotetext{e}{Cooling rate for the annealing.}
\end{deluxetable*}


\subsection{Comparison with ME0}\label{sec:compare_me0}

We apply the ME0 technique \citep{Metcalf1994ApJ...428..860M} to the same datasets as a baseline for comparison. ME0 minimizes a functional that combines the absolute divergence $|\nabla\cdot\mathbf{B}|$ and the vertical component of the electric current density $|J_z|$ across the entire domain. As the input is typically a 2D map of $\bm{B}$ assumed to be on a constant-$z$ plane, the term $\partial_z B_z$ needed for computing divergence is estimated from a reference potential field extrapolated with $B_z$ as the boundary condition \citep{Leka2009ASPC..415..365L}. A simulated annealing algorithm is then employed to decide whether to flip the azimuth angle of each pixel by $180^\circ$. For this work, we apply ME0 independently to each optical depth layer extracted from the MURaM simulations.

The parameters used in ME0 are listed in \autoref{tab:me0}. Most adopt default values, except for the \texttt{bthresh}, \texttt{athresh}, \texttt{neq}, and \texttt{tfactr}.
The first two parameters set the thresholds for transverse field strength, controlling the activation and deactivation of the numerical annealing algorithm. The third defines how many times a pixel is visited at a given ``temperature.'' The last determines the ``cooling rate'': the closer it is to $1$, the slower the cooling. We set \texttt{bthresh} and \texttt{athresh} as $1$ G for quiet Sun and plages, and $10$ G for the sunspot, and \texttt{neq} as $100$. The cooling rate is set to \texttt{tfactr}$=0.99$ for a more gradual annealing process. These updated parameters aim to improve ME0's results, particularly in weak field regions. As previously noted, we do not aim for a comprehensive comparison with ME0, but rather use it as a reasonable baseline.

The interpretation of the ME0 results can benefit from an empirically determined confidence mask, which marks regions where the ambiguity resolution is considered reliable. We construct such a confidence mask in \autoref{sec:me0conf} for a single optical depth layer at $\log_{10}\tau=-1$ for each case.

As discussed in Section~\ref{sec:test}, both HDD and ME0 perform reasonably where the transverse field is strong. The HDD method appears to excel in regions with weak to moderate field strengths, particularly for the quiet-Sun and the plage cases. The HDD solutions are also free of the checkerboard patterns in the final azimuth maps (\autoref{fig:5}): the more subtle checkerboard patterns in the intermediate predictions of $B_x$ and $B_y$ do not appear to affect the selection of the correct ambiguity resolution.
\ky{The checkerboard patterns observed in the intermediate predictions likely result from an uneven kernel overlap during the transposed-convolution upsampling, and can be mitigated by replacing it with a resize-convolution procedure \citep[i.e., explicit interpolation followed by a standard convolution,][]{odena2016deconvolution}. We will explore this avenue in future work.} 

We hypothesize that several different model designs between HDD and ME0 may contribute to the differences. First, while both HDD and ME0 encode divergence in their merit function, ME0 resorts to a reference potential field to estimate $\partial_z B_z$, which will differ from the multi-height informed values used in HDD. Extensions of ME0, such as ME0-Z \citep{Barnes2012AAS...22020609B}, can incorporate better estimates of $\partial_z B_z$ from other sources, but the availability of such data and processing is limited. Second, ME0 operates on individual $\tau$ layers under the assumption that each one is equivalent to a constant-$z$ surface. That assumption is only valid if the vertical fluctuation of $Z$ is small compared to the transverse spatial resolution, which is generally not valid for sub-arcsecond data as those used in this study. The added prediction of geometric height in HDD improves on this front. Third, high-resolution simulations suggest the existence of thin current layers. Intuitively, these structures are not the target solution of ME0 that include minimizing the total current content, which aims to ensure the smoothness across discontinuities. HDD on the other hand does not impose constraints on the electric currents.
We do not exclude ME0's capability to handle magnetograms in highly resolved structures, it is primarily designed for a constant-z layer and used as a downstream of Milne-Eddington inversions for HMI and Hinode SP pipelines (as mentioned in Sections \ref{sec:intro} and \ref{sec:test}). In this study, ME0 serves only as a baseline for our novel method, and its parameters are not fully optimized; a more suitable parameter set could potentially improve its results. However, such optimization is not the goal of this paper.


\subsection{Geometric Height}\label{sec:geo}

The HDD method successfully recovers many fine-scale variations in geometric height. Nevertheless, many are still missing, and the overall solution appears to be smoother than the ground truth. Moreover, the HDD results, including the geometric height, electric current density, and Lorentz force, generally agree better with the ground truth in lower layers, those below $\log_{10}\tau = -1.5$. These effects likely result from the specific form of the custom loss function (\autoref{eq:loss_div}): the fourth power in the numerator and squared magnetic field terms in the denominator emphasize stronger field regions, leading to better fitting in lower layers where the field strength is stronger, and smoother solutions in the sunspot compared to the quite-Sun or plage counterparts (\autoref{fig:10}(c)--(f)). We have carried out extensive explorations of alternative formulations for the loss functions; \autoref{eq:loss_div} currently performs best. Future work may improve on this front.

In previous works that estimate geometric heights based on 1D Stokes inversion results \citep{Puschmann2010ApJ...720.1417P,Loptien2018A&A...619A..42L,Loptien2020A&A...635A.202L}, the Wilson depression at each pixel was treated as a free parameter that minimizes a merit function. The approach involves high degrees of freedom (DOF) and suffers from the limitations of non-convex optimization techniques. \citet{Loptien2018A&A...619A..42L,Loptien2020A&A...635A.202L} considered the divergence-free condition in the Fourier space. They discarded high-frequency signals and consequently were expected to lose structural details. \ky{In addition, prior CNN-based work learned geometric heights directly from simulations, which showcased the capability of modern ML techniques on this specific task \citep{AsensioRamos2019A&A...626A.102A}.} Taking advantage of the physics law and modern ML technique, our approach can efficiently search for the ``global" minima in high-DOF optimization while preserving the fine-scale structural information.


Several recent works also tackle the problem of geometric heights associated with inversion results. For example, the FIRTEZ method integrates spectropolarimetric inversion with disambiguation, and iteratively calculates geometric heights using an MHS approximation \citep{Yabar2019A&A...629A..24P, Borrero2019A&A...632A.111B, Borrero2021A&A...647A.190B, Borrero2023A&A...669A.122B, Borrero2024A&A...687A.155B}. Its applicability to chromospheric observations is unclear, as magnetic fields inferred from non-LTE inversions and weak-field approximation methods tend to be stronger than those from the MHS model extrapolations, particularly for field strengths exceeding $300$~G \cite{Vissers2022A&A...662A..88V}. As another example, PINN techniques recently enabled the estimation of the chromosphere height relative to the photosphere through optimal fitting with a non-linear force-free field (NLFFF) model with multi-layer magnetograms as input \citep{Jarolim2024ApJ...963L..21J}. This approach assumes a flat photospheric plane and zero Lorentz force, whose validity remain questionable \citep{Metcalf1995ApJ...439..474M,Zhang2023ApJ...956L..17Z}. In contrast, our HDD method is designed with a minimal set of fundamental physical assumptions, and thus may have wider applications.
For example, previous studies \citep{Balthasar2018SoPh..293..120B} have reported systematic inconsistencies in the vertical magnetic field gradient between two common approaches: (1) estimating it from the field difference divided by an approximate spectral-line formation height, and (2) deriving it from the divergence-free condition using the horizontal partial derivatives of the magnetic field. While inaccurate height estimation is one possible cause, other factors may also contribute, which we do not address here. The HDD can provide height estimates that are consistent with the divergence-free condition, enabling a more accurate determination of magnetic gradients in the photospheric layers.


\section{Summary}\label{sec:summary}

In summary, our HDD method offers a new approach to reconstructing the 3D magnetic field in the lower solar atmosphere in regions with appreciable contributions to spectral line formation. The method leverages the new PIML tools and makes use of inversion results from spectropolarimetric observations. The key features of the HDD model are as follows:

\begin{enumerate}

\item HDD integrates physical laws into the powerful UNet3D framework. A flexible coordinate system and a set of custom loss functions facilitate accurate modeling.

\item HDD simultaneously resolves the azimuthal ambiguity and geometric height for a range of optical depths given an input estimated 3D magnetic field. This allows the reconstruction of a self-consistent 3D vector magnetic field throughout the lower solar atmosphere.

\item The preceding two features allow for physically meaningful calculations of downstream diagnostics, e.g., the 3D vector electric current density and Lorentz force distribution, among others.

\item Model training requires approximately $10^4$ iterations. While the process can be accelerated with multiple GPUs, it is still computationally expensive at the moment.

\end{enumerate}

A key advantage of our method is that it incorporates a physical law ($\nabla\cdot\bm{B}=0$) directly into the training process via a carefully designed custom loss function (\autoref{eq:loss}).
In minimizing the loss function, the model learns how unknown parameters (geometric height and azimuth) can be set such that the entire 3D system becomes consistent with the physical constraint.
This reduces dependence on a statistical brute-force pairing of large, labeled input-target datasets (as used in supervised approaches), greatly improves the model's ability to generalize to unseen observations, and directly addresses the limitation noted in the Introduction that \textit{``Models trained by MURaM simulations will only predict MURaM-like solutions."}
Consequently, the model is more adaptable and robust across diverse solar magnetic-field configurations, and provides a potential practical framework for pinpointing discrepancies between state-of-the-art radiative-MHD simulations and real observations.

Indeed, our method has been tested on simulated data for quiet Sun, plage, and sunspot regions, with excellent accuracy. Additional tests (\autoref{sec:a1} and \ref{sec:a2}) hints at its capability when applied to unseen and more realistic (noisy, lower-resolution) input. The free parameters, e.g., the weighting for various loss function terms, can be adjusted according to optimize performance across diverse datasets. In addition, the coordinate system used in this paper can self-adjust geometric height to handle off-disc-center case, i.e., the tilt and depression of the constant-$\tau$ surface will be determined by the predicted $Z$ directly. But we have not tested this capability. We plan to demonstrate its off-disc-center performance and incorporate the temporal dimension in the near future, enabling application to various disc positions and time series observations.


\begin{acknowledgments}
This work is supported by NSF/AAG award \#2008344. X.~Sun is additionally supported by NSF CAREER award \#1848250 and the state of Hawai`i. J.~Liu acknowledges the support of the DKIST Ambassador program. This material is based upon work supported by the NSF National Center for Atmospheric Research, which is a major facility sponsored by the U.S. National Science Foundation under Cooperative Agreement No. 1852977. We thank G. Barnes and especially acknowledge K. D. Leka for valuable input. We would like to acknowledge the GPU cluster at the Institute for Astronomy, University of Hawai`i at M\=anoa, and the high-performance computing support from \textit{Cheyenne} (doi: \href{https://dx.doi.org/10.5065/D6RX99HX}{10.5065/D6RX99HX}) provided by NCAR's Computational and Information Systems Laboratory, sponsored by the NSF. 
The technical support and advanced computing resources from UH Information Technology Services -- Cyberinfrastructure, funded in part by the NSF CC* awards \#2201428 and \#2232862 are gratefully acknowledged.
\end{acknowledgments}


\software{\texttt{PyTorch} \citep{Paszke2019PyTorchAI}, \texttt{Matplotlib} \citep{Hunter2007}, \texttt{SciPy} \citep{2020SciPy-NMeth}, \texttt{Numpy} \citep{harris2020array}, and \texttt{pytorch-3dunet} \citep{Wolny10.7554/eLife.57613}.}



\begin{figure*}[t!]
\centering
\fig{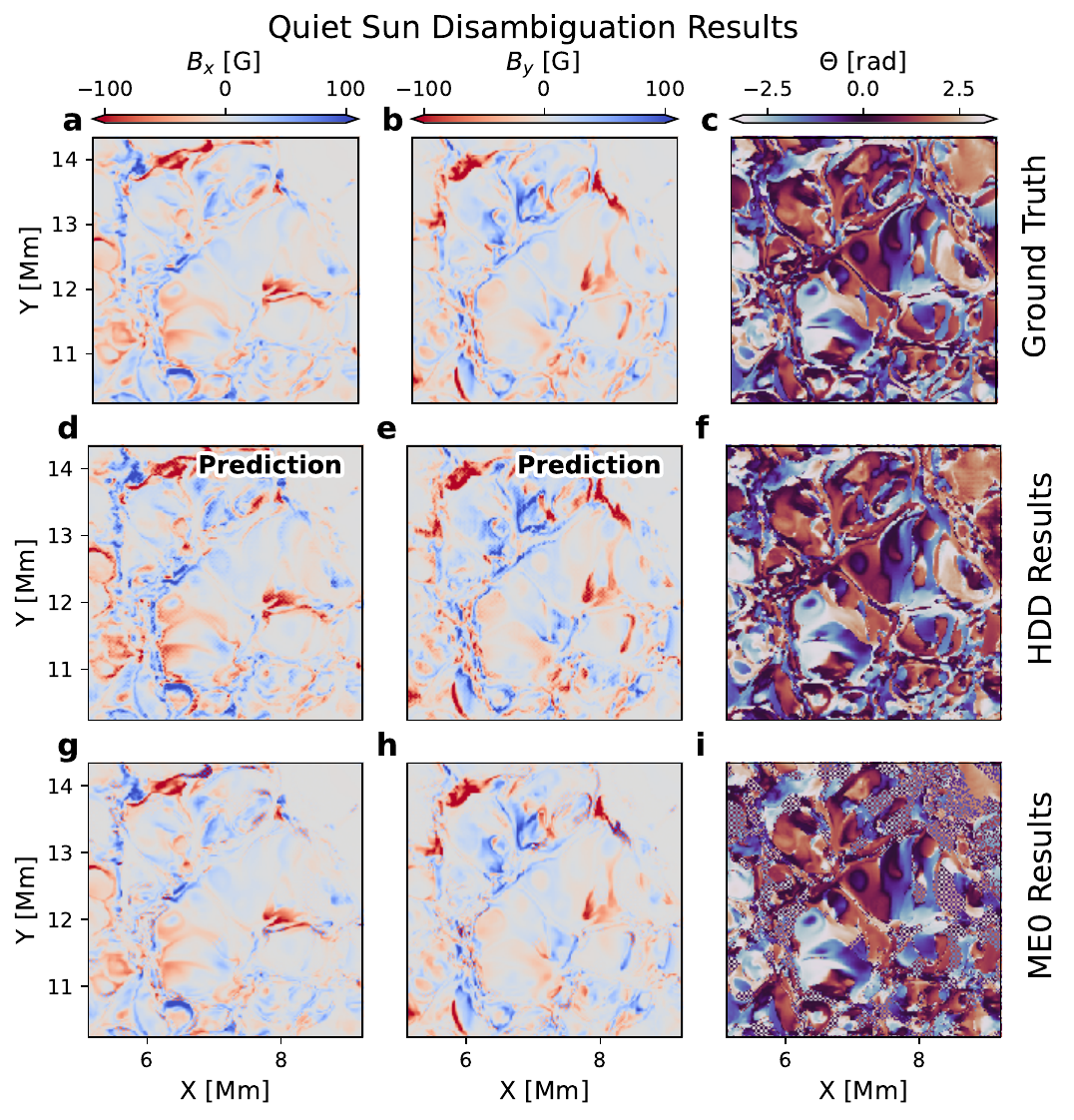}{0.59\textwidth}{}
\caption{Same as \autoref{fig:5}, but for the quiet Sun dataset shown in \autoref{fig:3}(d). \label{fig:4}}
\vspace{5mm}

\fig{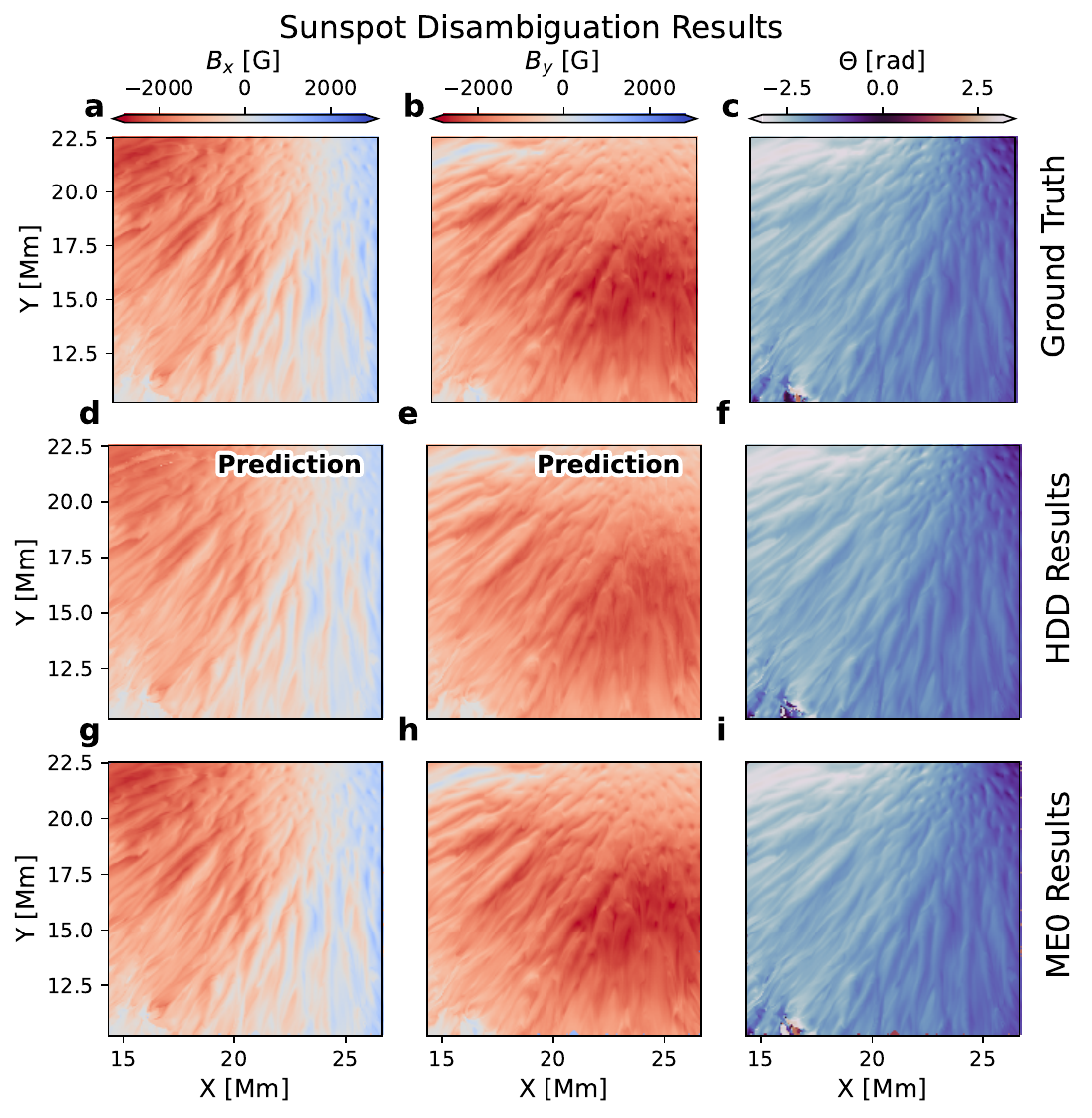}{0.59\textwidth}{}
\caption{Ssame as \autoref{fig:5} but for the sunspot dataset that shown in \autoref{fig:3}(f). \label{fig:6}}
\end{figure*}


\appendix

\section{Test Results for Quiet Sun and Sunspot Simulations}\label{sec:a0}
\begin{figure*}[t!]
\centering
\fig{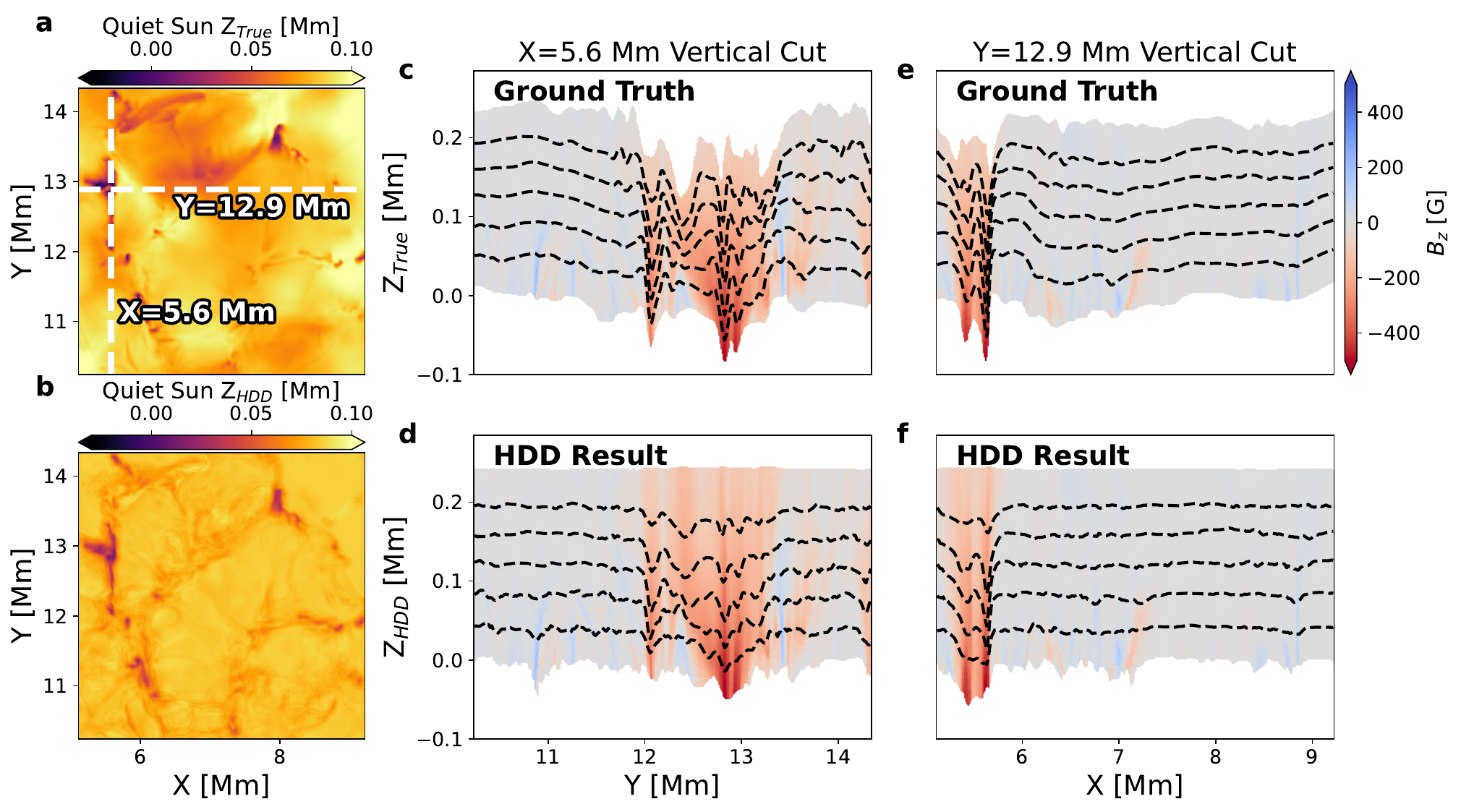}{0.76\textwidth}{}
\caption{Same as \autoref{fig:9} but for the quiet sun results. \label{fig:8}}
\vspace{5mm}

\centering
\fig{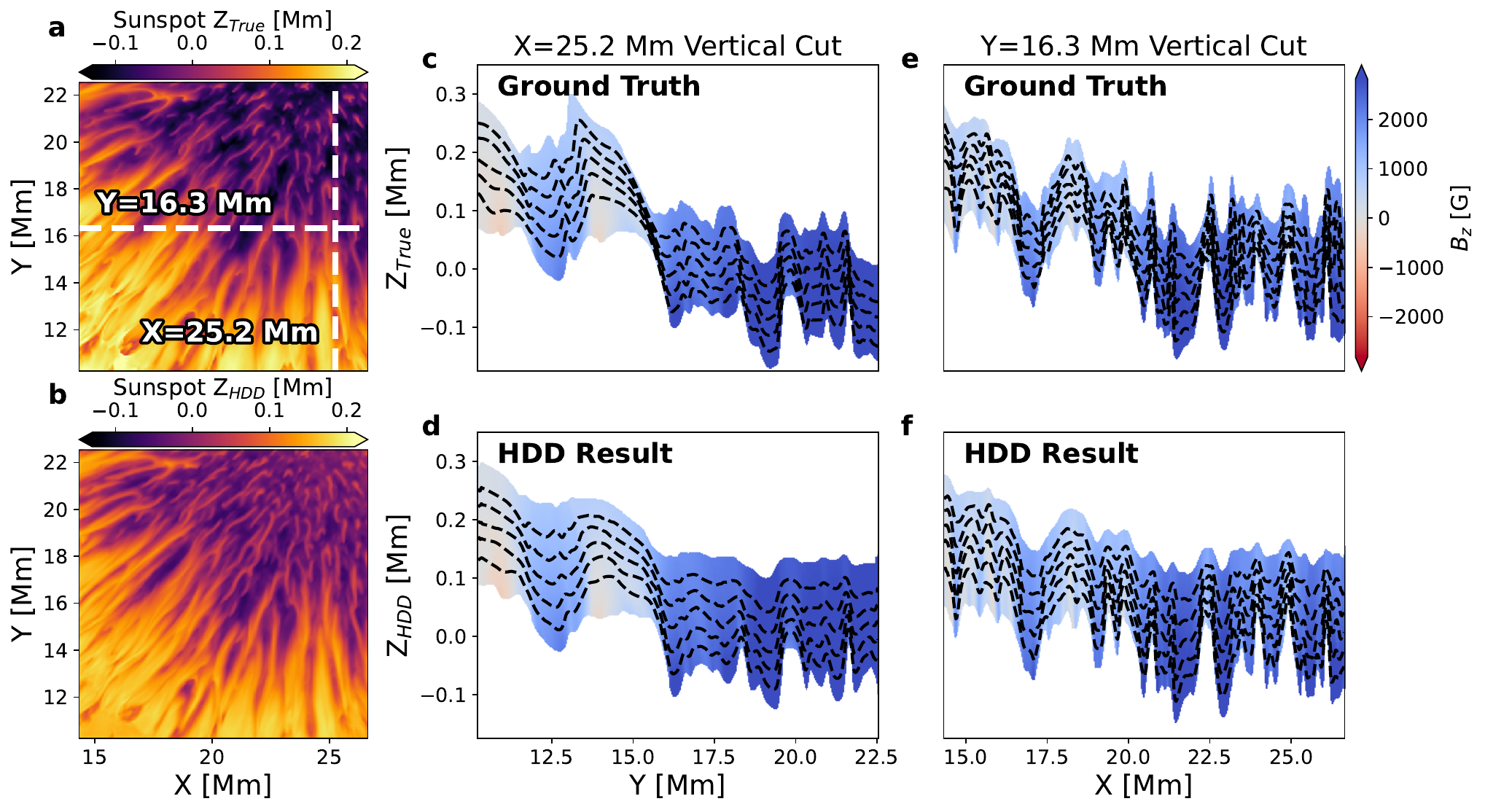}{0.76\textwidth}{}
\caption{Same as \autoref{fig:9} but for the sunspot results.\label{fig:10}}
\end{figure*}

\begin{figure*}[t!]
\centering
\fig{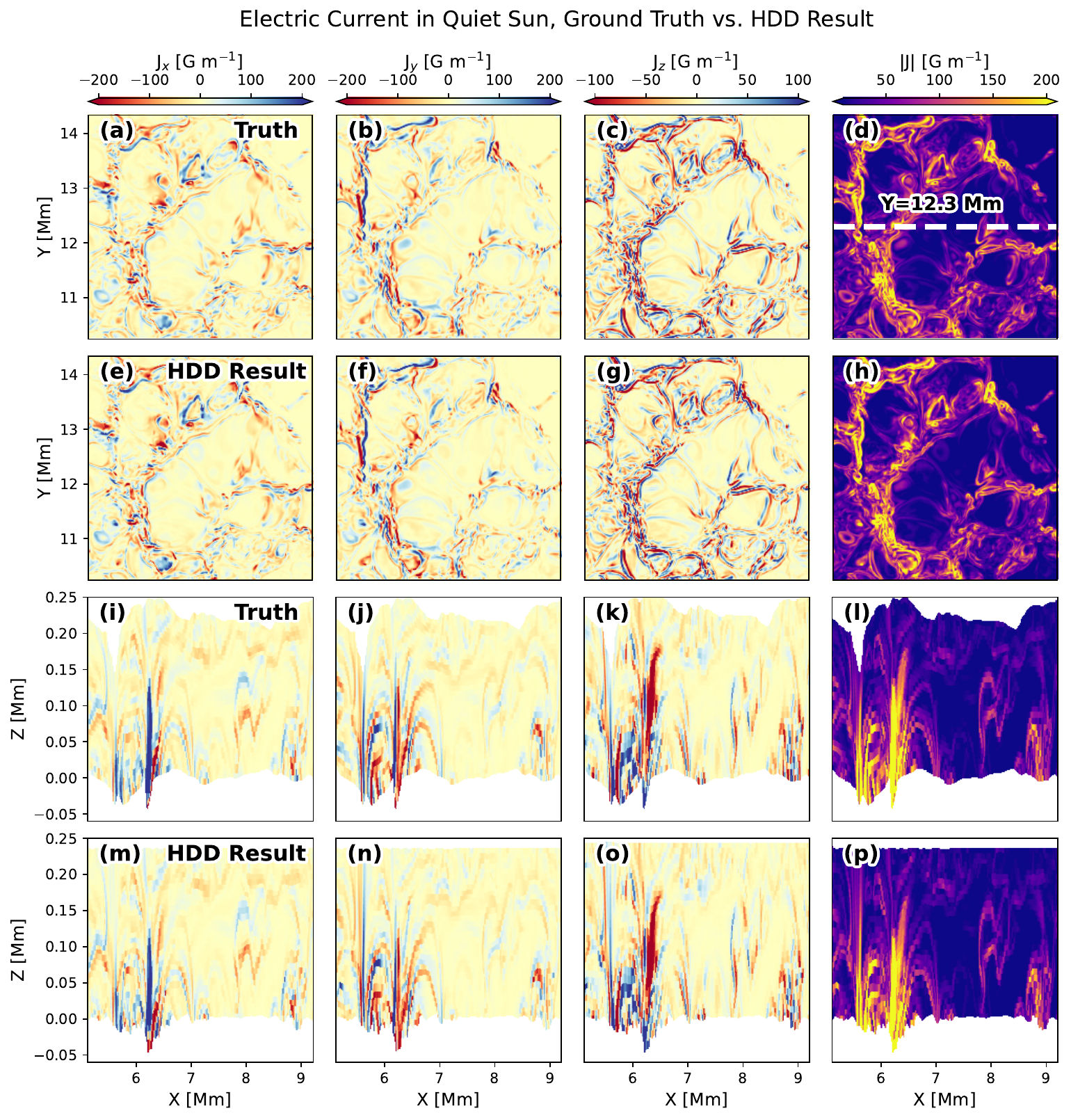}{0.75\textwidth}{}
\caption{Same as \autoref{fig:13} but for the quiet sun results. \label{fig:12}}
\end{figure*}

\begin{figure*}
\centering
\fig{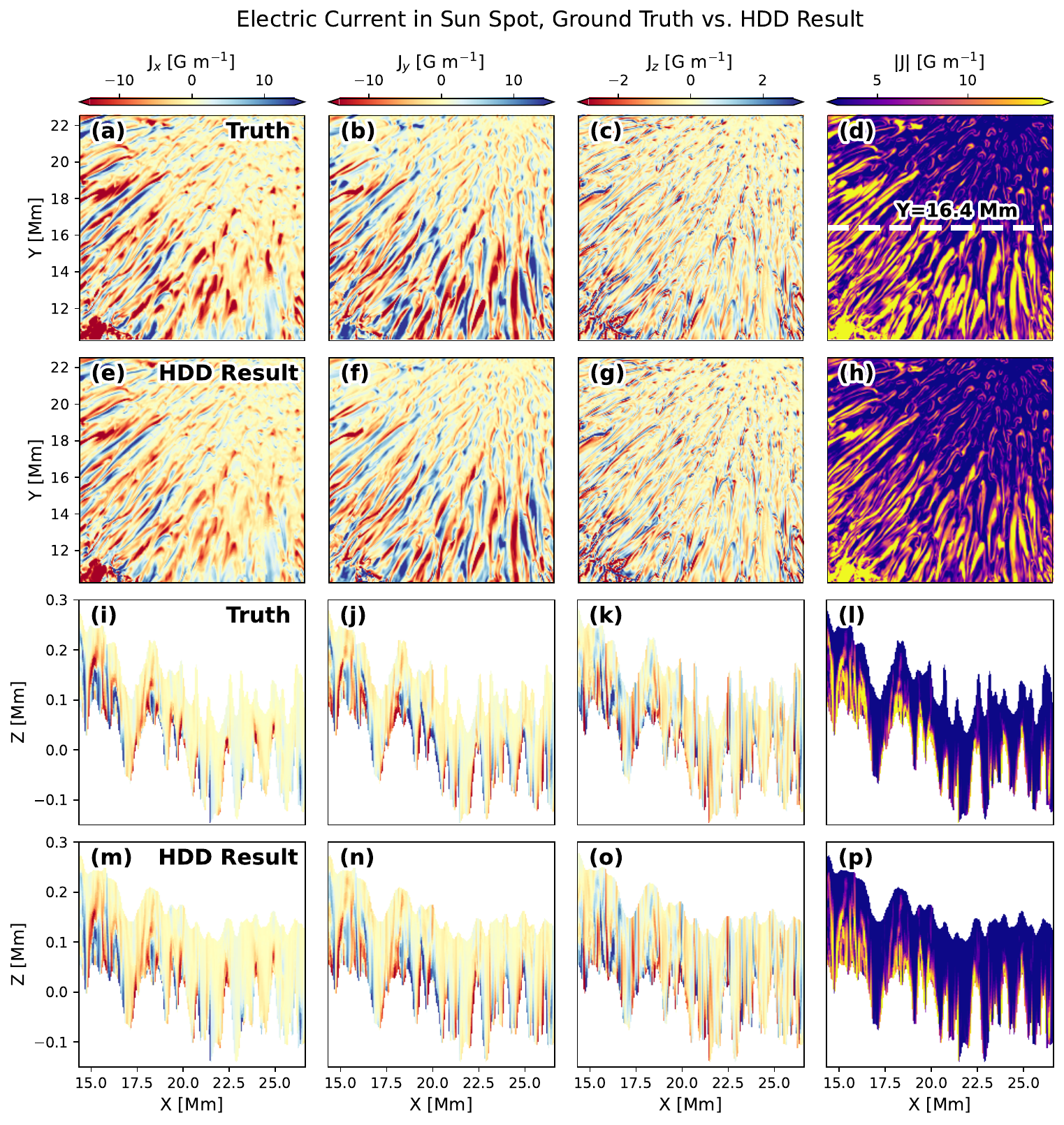}{0.75\textwidth}{}
\caption{Same as \autoref{fig:13} but for the sunspot results. \label{fig:14}}
\end{figure*}

\begin{figure*}
\centering
\fig{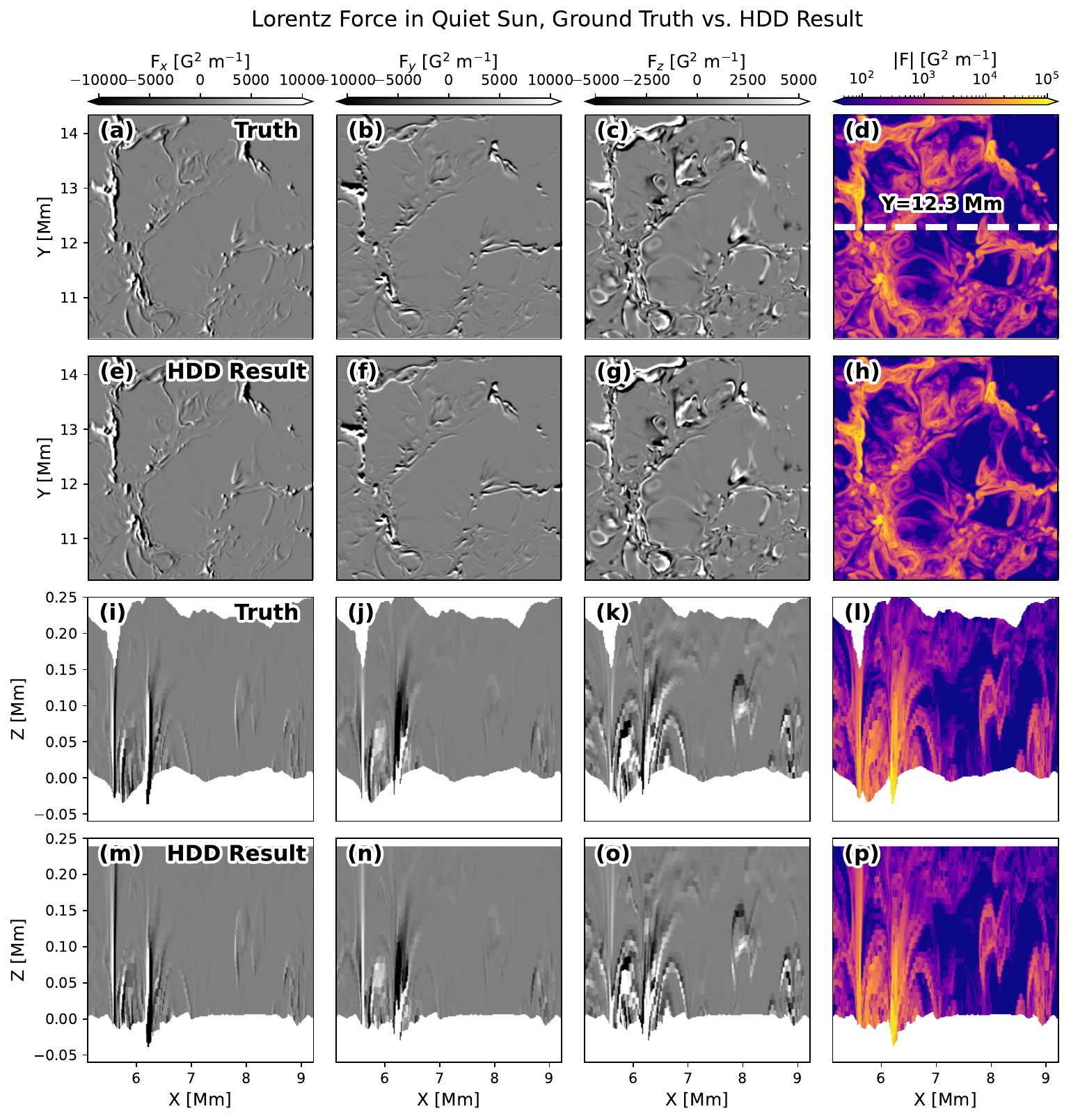}{0.75\textwidth}{}
\caption{The same as \autoref{fig:12} but showing the Lorentz force. \label{fig:17}}
\end{figure*}

\begin{figure*}
\centering
\fig{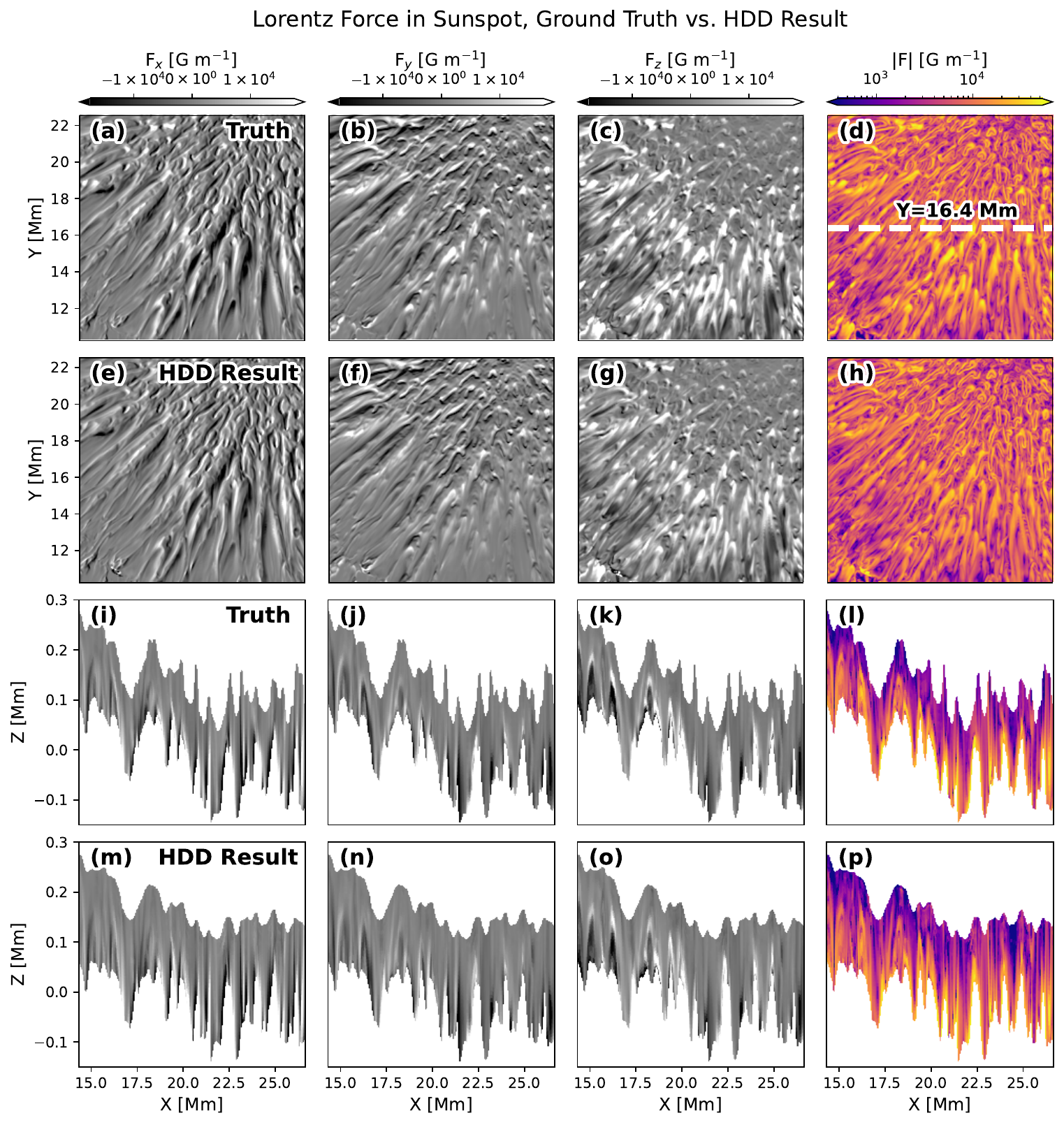}{0.75\textwidth}{}
\caption{The same as \autoref{fig:14} but showing the Lorentz force. \label{fig:19}}
\end{figure*}

\begin{figure*}
\centering
\fig{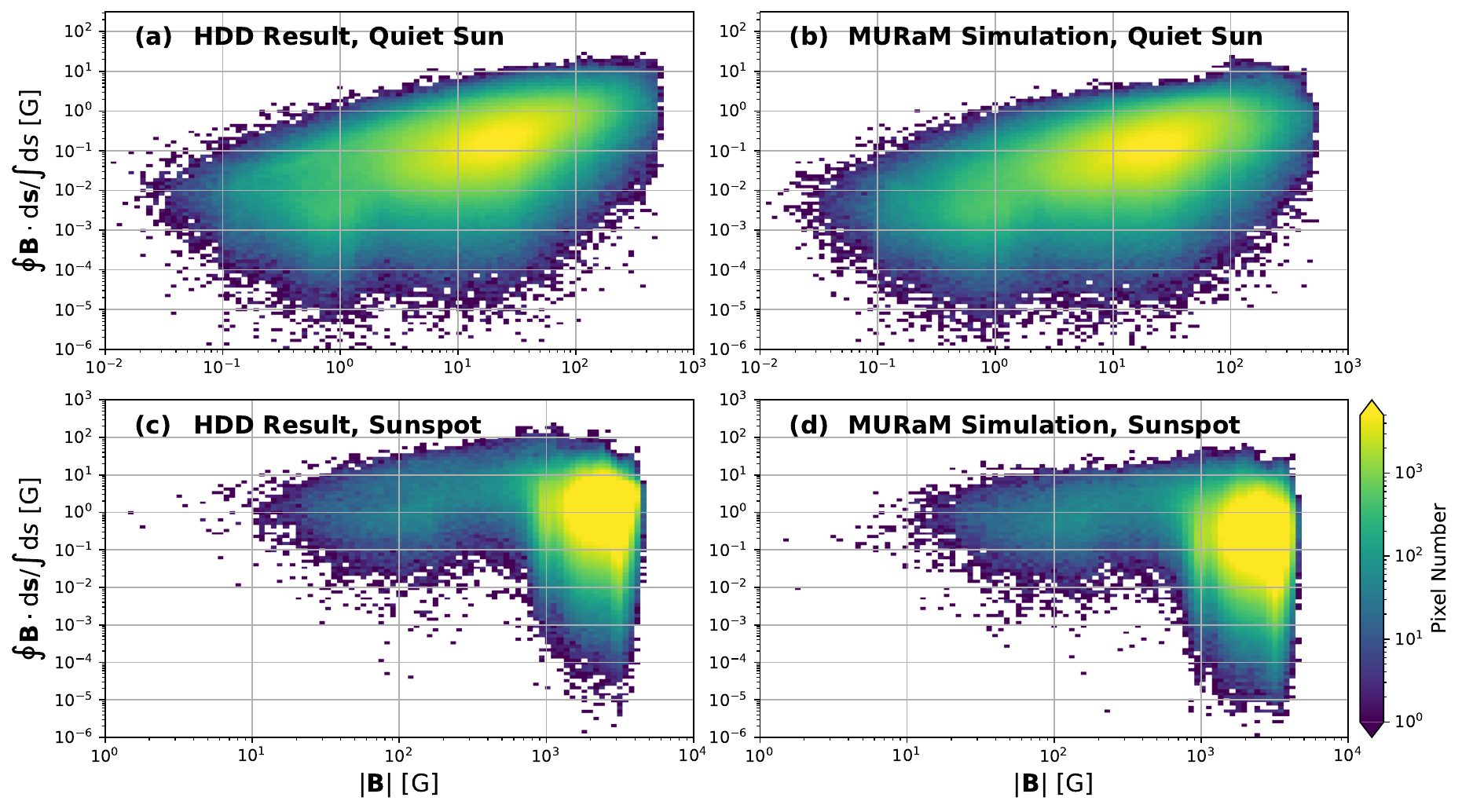}{0.75\textwidth}{}
\caption{Same as \autoref{fig:divb}, but for the quiet Sun and sunspot. Panels (a) and (b) show the results for the quiet Sun, while panels (c) and (d) correspond to the sunspot.} \label{fig:divb_qs_ss}
\end{figure*}

The results presented here directly follow those for the Plage simulation described in the main text. 
The disambiguation results for the quiet sun and sunspot simulations are shown in \autoref{fig:4} and \autoref{fig:6}, following the same format as \autoref{fig:5}. The predicted geometric heights are shown in \autoref{fig:8} and \autoref{fig:10}. Comparisons between the ground truth and HDD-derived values of the electric current are presented in \autoref{fig:12} and \autoref{fig:14}, and those for the Lorentz force are presented in \autoref{fig:17} and \autoref{fig:19}, respectively. The 2D histograms of the residual magnetic flux across cell surfaces versus the average transverse magnetic field within the corresponding cell for the quiet Sun and sunspot simulations are shown in \autoref{fig:divb_qs_ss}. For the HDD solutions, the median values of integrated divergence errors in the finite volume calculation of the HDD grid are $0.12$~G for the quiet Sun and $1.3$~G for the sunspot, compared to $0.08$~G and $0.19$~G, respectively, from the MURaM simulation.


\section{Model Performance on Unseen Data}\label{sec:a1}

\begin{figure*}
\centering
\fig{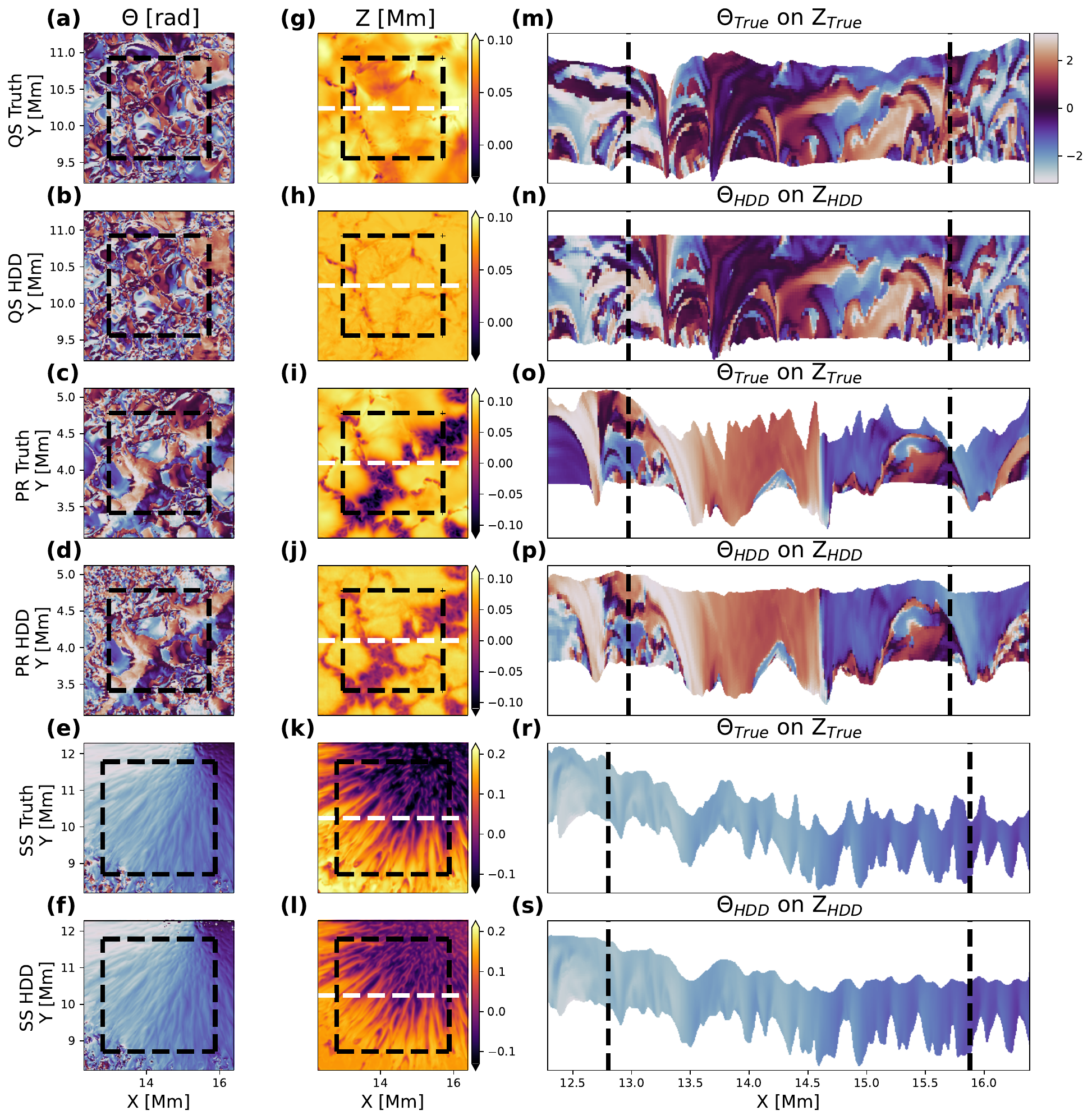}{0.85\textwidth}{}
\caption{Columns from left to right show the predicted azimuth angle on $\log_{10}\tau=-1$ surfaces ((a)--(f)), geometric height on $\log_{10}\tau=-1$ surfaces ((g)--(l)), and angle distribution along vertical cuts ((m)--(s)), respectively. Rows from top to bottom display the ground truth and predictions for the quiet Sun (QS), plage region (PR), and sunspot (SS), respectively. Black dashed lines enclose the regions used for model training, while the white dashed line marks the location of the vertical slice. \label{fig:a1}}
\end{figure*}

The HDD model can be regarded as a learnable PDE solver capable of learning patterns from data and using them to solve for the magnetic field based on the divergence-free equation. To evaluate its learning capabilities, we apply the model to a dataset that it has not seen previously. The results are presented in \autoref{fig:a1}. The training data is the portion within the black dashed boxes. The portions on the periphery represents the new data handed directly to the trained network. 

The predicted azimuth angle and geometric height are compared with the ground truth. The results demonstrate that the HDD model can provide reasonable results for the untrained data. This highlights the model's capacity to generalize beyond its training domain.


\section{Confidence Mask for ME0}\label{sec:me0conf}

Confidence masks in ME0 are computed through 100 independent repetitions, each initiated with a different $\texttt{iseed}$ (from 1 to 100) to initialize the random number generator. The confidence value is determined by the occurrence rate of the majority solution. That is, if $90$ runs indicate that the azimuth should be flipped (or not) for a pixel, the confidence mask will adopt a $90\%$ value there. Given the constraint on computing resource, we generate the confidence mask only for one optical depth layer at $\log_{10}\tau=-1$ for MHD case. The metric values presented in \autoref{tab:compare} are computed within the $90\%$ confidence mask.


\section{Model Performance on Noisy Data}\label{sec:a2}

To evaluate the performance of the HDD code under more realistic conditions, we conduct tests using data binned to lower resolution and with added noise. Gaussian noise with a standard deviation of $10$~G is added to the quiet Sun and plage region datasets, and $100$~G to the sunspot dataset. The original data are then rebinned by $2\times 2$ to mimic the spatial resolution of DKIST, about ${\rm d}x=32$ km for the quiet Sun and plage, and ${\rm d}x=64$ km for the sunspot regions. 
A 3D median filter with a kernel size of $3\times3\times3$ vertices was applied before running the HDD code on the dataset, as typically done to denoise observational results before further analysis.
We note that this noise test is a highly simplified version, and does not follow the \textit{ab initio} approach of simulating noisy, rebinned spectral lines and inverting them, as done in \citet{Leka2012SoPh..277...89L}.

In order to handle the uncertainties of the transverse field, we modify the loss functions for the magnetic field components (\autoref{eq:bt} and \autoref{eq:parallel}).
\begin{equation}
    {\rm loss}_{B_t}=\left<\frac{\max\{(B_{x,{\rm pred}}^2+B_{y,{\rm pred}}^2 - B_{x,{\rm in}}^2-B_{y,{\rm in}}^2)^2-B_{t,{\rm err}}^4,0\}}{B_{x,{\rm in}}^2+B_{y,{\rm in}}^2+\epsilon}\right>,
\end{equation}
\begin{equation}
  {\rm loss}_{\rm parallel}=\left<\frac{\max\{(B_{x,{\rm pred}}B_{y,{\rm in}} - B_{y,{\rm pred}}B_{x,{\rm in}})^2-E_{\rm parallel},0\}}{|B_{\rm in}|^2+\epsilon}\right>,
\end{equation}
where $B_{t,{\rm err}}$ is the uncertainty in the transverse field, i.e., $10$ G for the quiet Sun and plages, and $100$ G for the sunspot. The error of the parallelism is formulated as $E_{\rm parallel}=\max\{E_{{\rm parallel},1}, E_{\rm parallel,2}\}$, where
\begin{equation*}
\begin{split}
E_{\rm parallel,1} &= ((B_{x,{\rm in}}+B_{t,{\rm err}})B_{y,{\rm in}} - (B_{y,{\rm in}}-B_{t,{\rm err}})B_{x,{\rm in}})^2,\\
E_{\rm parallel,2} &= ((B_{x,{\rm in}}-B_{t,{\rm err}})B_{y,{\rm in}} - (B_{y,{\rm in}}+B_{t,{\rm err}})B_{x,{\rm in}})^2.
\end{split}
\end{equation*}
For comparison, the ME0 parameters \texttt{athresh} and \texttt{bthresh} shown in \autoref{tab:me0} are set to $10$ for quiet Sun and plage regions, and $100$ for sunspots.

\begin{figure*}
\centering
\fig{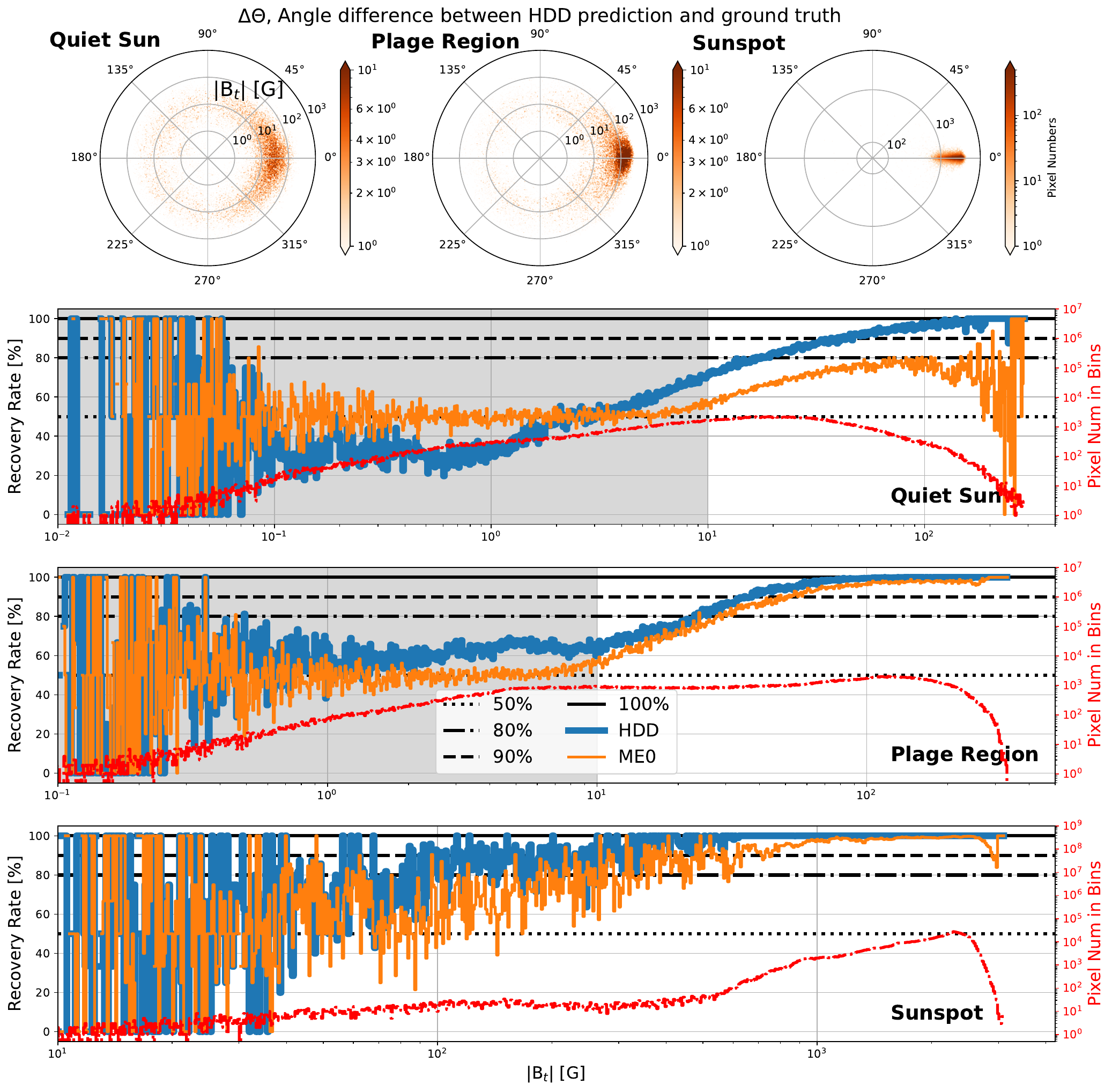}{0.75\textwidth}{}
\caption{Same as \autoref{fig:7} but for the results from the noisy, rebinned dataset. \label{fig:a2}}
\end{figure*}

The recovery rates for data with added noise are illustrated in \autoref{fig:a2}. The performance for azimuth prediction is reasonable when the transverse field is strong, but suffers significantly in weaker field regions. This shows up as a diffuse distribution of the angle difference $\Delta\Theta$ compared to \autoref{fig:11}. At a noise level of $10$~G, the recovery rate in both the quiet Sun and plages remains above $80$~\% for $|B_t|>20$~G. In this range, the HDD and ME0 methods show similar performance in the plage case, while the HDD method performs better in the quiet Sun case. In the sunspot case, where the noise level reaches approximately $100$~G, the HDD results reach a recovery rate above $80$ \% for approximately $|B_t|>300$~G. 

\begin{figure*}
\centering
\fig{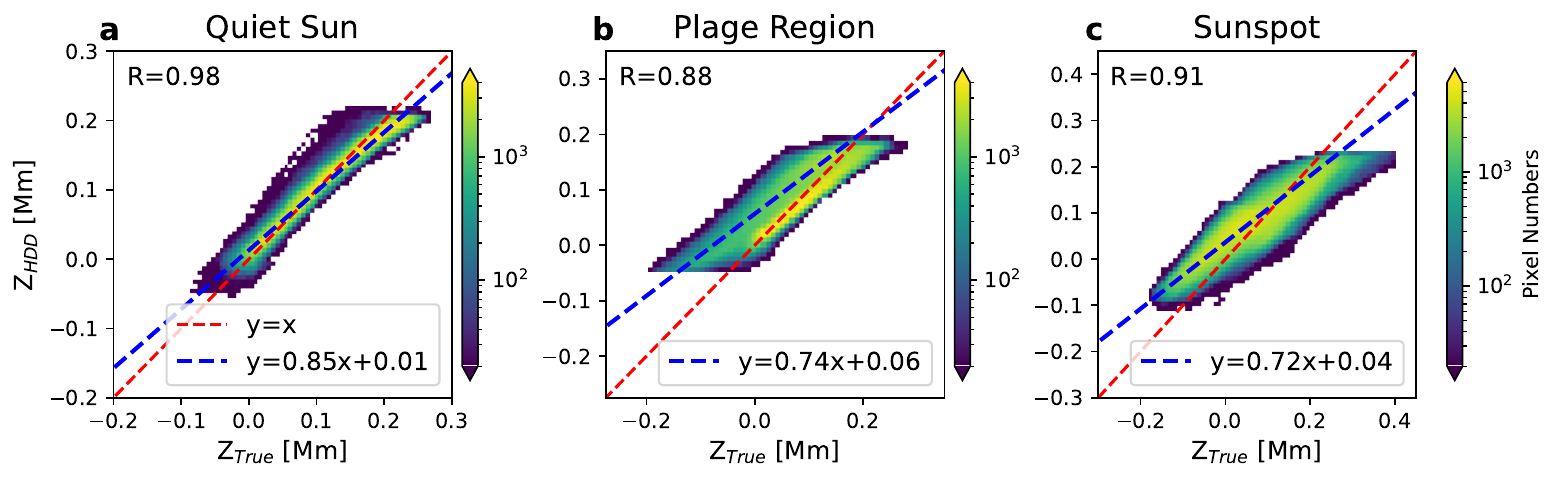}{0.8\textwidth}{}
\caption{The Same as \autoref{fig:11} but for the results from the noisy, rebinned dataset. \label{fig:a3}}
\end{figure*}

\begin{figure*}
\centering
\fig{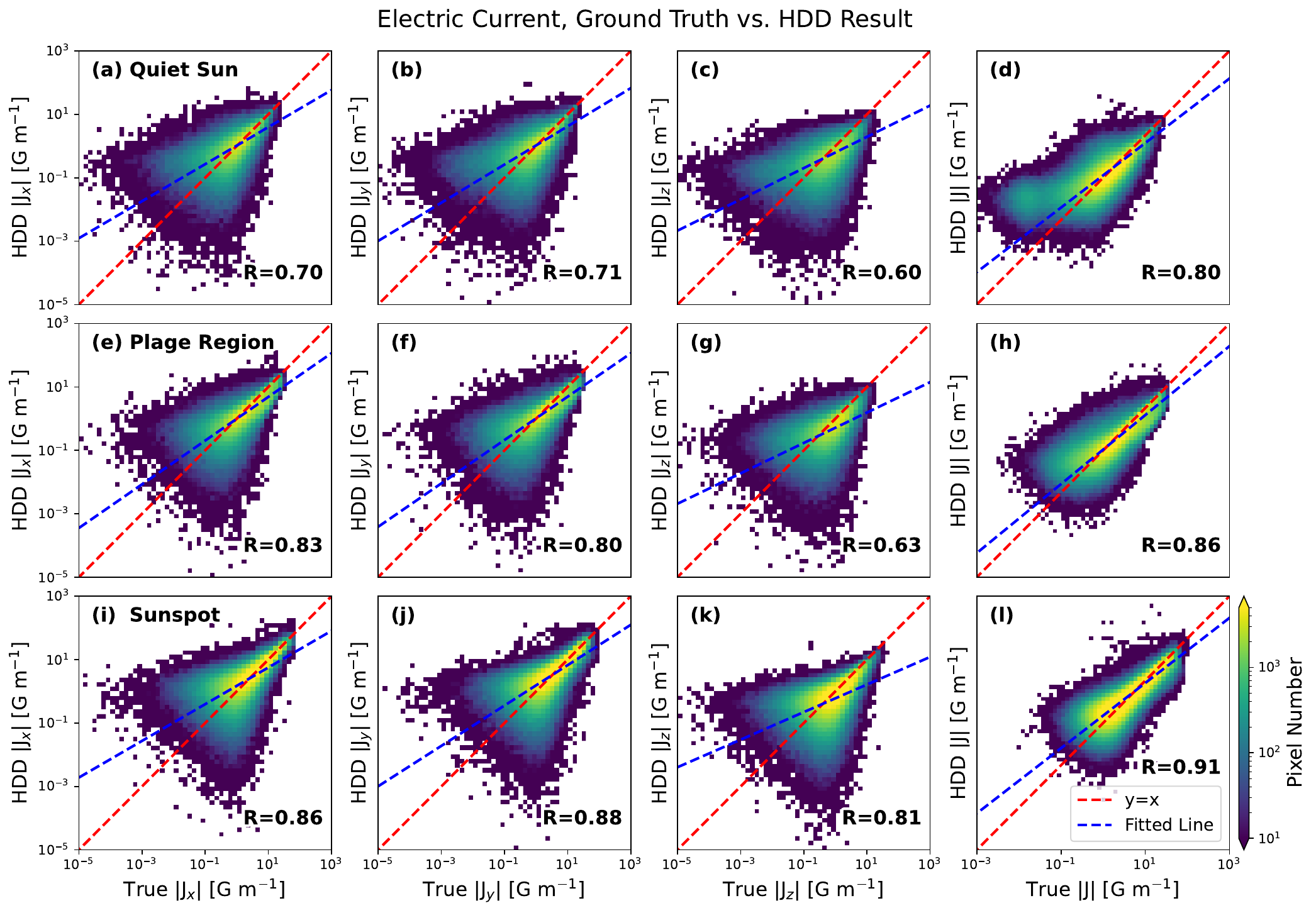}{0.8\textwidth}{}
\caption{The Same as \autoref{fig:15} but for the results from the noisy, rebinned dataset. \label{fig:a4}}
\end{figure*}

The results for geometric heights, electric current density, and Lorentz force are shown shown in \autoref{fig:a3}, \autoref{fig:a4}, and \autoref{fig:a4} respectively. While the Pearson correlation coefficients are slightly lower than those for the noiseless data, they are still reasonable overall, and appear to be at least on par with existing methods \citep{PastorYabar2021AA...656L..20P,Borrero2023AA...669A.122B}. These tests demonstrate the reliability of the HDD method and its applicability to real observations.


\begin{figure*}
\centering
\fig{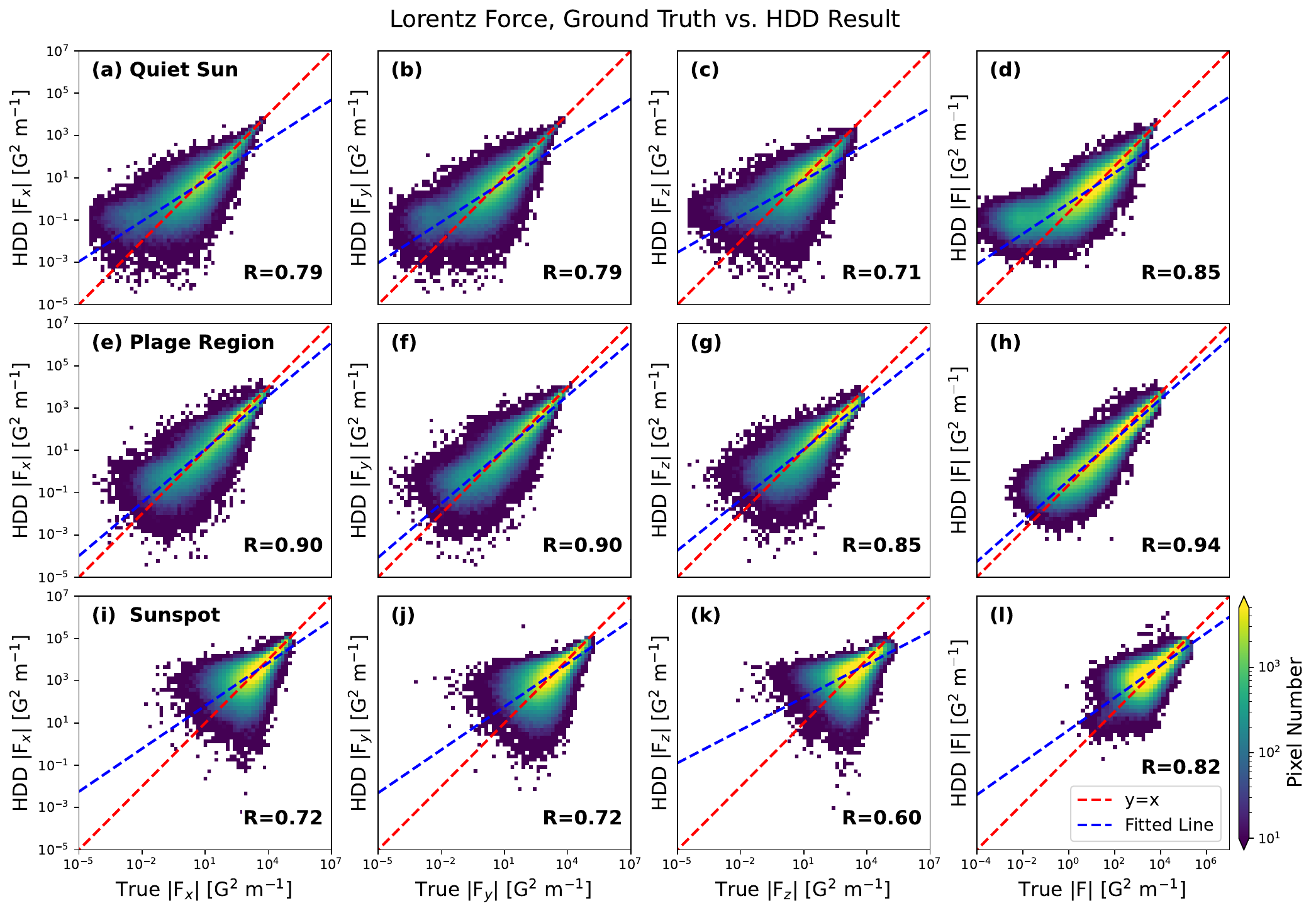}{0.8\textwidth}{}
\caption{The Same as \autoref{fig:20} but for the results from the noisy rebinned dataset. \label{fig:a5}}
\end{figure*}

\end{CJK*}

\clearpage


\bibliographystyle{aasjournal}


\end{document}